\documentclass[12pt,a4paper]{article} 

\usepackage{graphics,graphicx,placeins,afterpage,makecell,tablefootnote,
multicol,multirow,caption,textcomp,lineno,tabularx,subfigure,booktabs,
courier,collref,setspace,float}
\usepackage[utf8]{inputenc}
\usepackage[english]{babel}
\usepackage[backend=bibtex,style=authoryear,bibstyle=authoryear,maxcitenames=2,maxbibnames=7,useprefix=true]{biblatex}

\DeclareNameAlias{sortname}{last-first}

\begin{document}
\pagestyle{plain}
\pagenumbering{arabic}
 
 %%%%% Title %%%%% 
\begin{center}
{\LARGE{Abundance Measurements of Titan's Stratospheric HCN,
      HC$_3$N, C$_3$H$_4$, and CH$_3$CN from ALMA Observations}}
\end{center}
\bigskip

 %%%%% Authors %%%%% 
\begin{center}
Authors: Alexander E. Thelen$^a$\footnote{Corresponding
  author. (A. E. Thelen) Email
  address: athelen@nmsu.edu. Postal Address:
  Department of Astronomy, New Mexico State University, PO BOX 30001,
  MSC 4500, Las Cruces, NM 88003-8001.} %Telephone: (831)-332-1899.}
, C. A. Nixon$^b$,
N. J. Chanover$^a$, M. A. Cordiner$^{b,c}$, E. M. Molter$^d$, 
N. A. Teanby$^e$, P. G. J. Irwin$^f$, J. Serigano$^g$, S. B. Charnley$^b$
\bigskip

{\footnotesize{$^a$New Mexico State University, $^b$NASA Goddard Space
    Flight Center, $^c$Catholic University of
America, $^d$University of California, Berkeley, $^e$University of Bristol,
$^f$University of Oxford, $^g$Johns Hopkins University}}
\end{center}
\bigskip

 %%%%% Abstract %%%%% 
\begin{abstract}
Previous investigations have employed more than 100 close observations
of Titan by the \textit{Cassini}
orbiter to elucidate connections between the
production and distribution of Titan's
vast, organic-rich chemical inventory and its atmospheric dynamics. 
However, as Titan transitions into northern summer, the lack of incoming data
from the \textit{Cassini} orbiter presents a potential barrier to the continued
study of seasonal changes in Titan's atmosphere. In our
previous work (Thelen, A. E. et al. [2018]. Icarus 307, 380--390), we
demonstrated that the Atacama Large Millimeter/submillimeter Array
(ALMA) is well suited for measurements of Titan's atmosphere
in the stratosphere and lower mesosphere ($\sim100-500$
km) through the use of spatially resolved (beam sizes
$\textless1''$) flux calibration
observations of Titan. Here, we derive vertical abundance profiles of
four of Titan's trace atmospheric species from the same 3 independent spatial
regions across Titan's disk during the same epoch (2012 to 2015): HCN, HC$_3$N, C$_3$H$_4$,
and CH$_3$CN. We find that Titan's minor
constituents exhibit large latitudinal variations, with enhanced
abundances at high latitudes compared to equatorial measurements; this
includes CH$_3$CN, which eluded previous detection by \textit{Cassini} in
the stratosphere, and thus spatially resolved abundance measurements were
unattainable. Even over the short 3-year period, vertical profiles and
integrated emission maps of these molecules allow
us to observe temporal changes in Titan's atmospheric circulation during
northern spring. Our derived abundance profiles are comparable to contemporary
measurements from \textit{Cassini} infrared observations, and we find
additional evidence for subsidence of enriched air onto Titan's south
pole during this time period. Continued observations of Titan with
ALMA beyond the summer solstice will enable further study of how
Titan’s atmospheric composition and dynamics respond to seasonal changes.
\end{abstract}
\bigskip

 %%%%% Keywords %%%%% 
\begin{center}
\end{center}
\bigskip

 %%%%% Paper Body %%%%% 

\section{Introduction} \label{introduction}
%\begin{linenumbers}
Saturn's largest moon, Titan, is host to a dense, dynamic atmosphere
rich with trace organic molecules produced through N$_2$ and CH$_4$
generated photochemistry. Many hydrocarbon (C$_X$H$_Y$) and nitrile
(C$_X$H$_Y$[CN]$_Z$) species have been detected throughout Titan's
atmosphere, often showing vertical gradients from their primary
formation site in the upper atmosphere (upwards of
700 km) through N$_2$/CH$_4$ dissociation and ionospheric
interactions, to condensation near the tropopause at altitudes
below 80 km (see review by \cite{horst_17}).

Decades after the initial discovery of Titan's atmosphere through the
spectroscopic detection of CH$_4$ \parencite{kuiper_44}, additional
trace hydrocarbons C$_2$H$_2$, C$_2$H$_4$, and C$_2$H$_6$ were
detected through ground-based observations in the
IR \parencite{gillett_73, gillett_75}. N$_2$ and Titan's most abundant
nitriles -- HCN (\textit{hydrogen cyanide}), HC$_3$N (\textit{cyanoacetylene}), and C$_2$N$_2$ -- were
discovered through \textit{Voyager 1} Ultraviolet Spectrometer
(UVS) and Infrared Spectrometer (IRIS)
observations during the spacecraft's flyby of Titan in
1980 \parencite{broadfoot_81, hanel_81, kunde_81}, in addition to more
complex hydrocarbons such as C$_3$H$_8$ and
C$_3$H$_4$ (or CH$_3$CCH, \textit{methylacetylene}; \cite{maguire_81}). Many of these trace constituents
were found to be enhanced at mid to high northern latitudes ($\textgreater50^\circ$N, then in
winter) compared
to the equator and southern latitudes. In 
particular, the nitriles were enhanced by up to an order of
magnitude near the north pole. This enrichment was initially attributed to
the shielding of the winter pole from UV
radiation due to Titan's obliquity of $\sim26^\circ$, mitigating the rapid depletion of nitriles and some
hydrocarbons in the stratosphere, and potential seasonal effects \parencite{yung_87, coustenis_95}.

Upon the arrival of the \textit{Cassini} orbiter to the Saturnian system
in 2004 (nearly one Titanian year after the \textit{Voyager 1}
flyby),
more in-depth observations of Titan's atmospheric composition and dynamics were
possible through the close monitoring of the moon over 127 targeted
flybys, often within 1000 km of the moon’s surface and well within
its ionosphere, and the deployment of the \textit{Huygens} probe to
Titan's surface in 2005. A campaign of studies investigating Titan's
atmosphere revealed further complex chemistry, a wealth
of unidentified heavy positive and negative ions in the upper atmosphere, additional
hydrocarbons and nitriles, and
confirmation that the distributions of Titan's complex chemical species
are connected to its atmospheric dynamics (see reviews in
\cite{bezard_14, vuitton_14}). The enhancement
of many trace chemicals above Titan's north pole was again observed
during northern winter, followed by a change from
south-to-north circulation to a two cell pattern with upwelling onto
both poles, and finally into a completely reversed, north-to-south
circulation cell in 2011 \parencite{flasar_05, teanby_08b, teanby_10a, teanby_12,
  vinatier_15, coustenis_16}.

The results of \textit{Voyager 1} studies of Titan prompted the
use of mm/sub-mm ground-based observations \parencite{paubert_84}, leading
to the confirmation of the existence of HCN, HC$_3$N \parencite{paubert_87, bezard_92}, 
and the detection of CH$_3$CN (\textit{methyl cyanide}) \parencite{bezard_93} with the IRAM 30-m
telescope; the latter
molecule appeared to be of comparable abundance to Titan's other
nitriles through laboratory experiments \parencite{raulin_82}, but
eluded detection in the IR during the \textit{Voyager} and
\textit{Cassini} eras. The vertical profiles of these molecules
have since been studied from the ground with IRAM \parencite{marten_02}, the Submillimeter
Array (SMA; \cite{gurwell_04}), and most recently, the Atacama Large
Millimeter/submillimeter Array (ALMA; \cite{cordiner_14,
  molter_16}), which is also capable of detecting C$_3$H$_4$, HNC,
C$_2$H$_3$CN, C$_2$H$_5$CN, and potentially other
trace nitriles \parencite{teanby_18, cordiner_15, palmer_17, lai_17}. While early mm/sub-mm studies of Titan have
resulted in disk-averaged measurements of these minor constituents,
ALMA currently provides the capabilities to study the spatial variation of many
species through resolved observations of Titan, which is $\sim1''$ on the sky
(including its extended atmosphere) compared to the maximum resolution
obtainable with ALMA of few--10s of mas. The frequent observations of Titan
for ALMA flux calibration measurements facilitates the continuation of
\textit{Cassini's} legacy, allowing for studies of Titan's climate and
atmospheric chemistry beyond the northern summer solstice.  

\textcite{thelen_18} -- hereafter referred to as `Paper I' -- showed that ALMA flux calibration
observations of Titan enable the measurement of spatial
variations in stratospheric temperature. While the viewing geometry
and spatial resolution of early flux calibration data only permitted
large latitudinal averages on three separate regions on Titan, the
temperature measurements discussed in
Paper I were in agreement with those found by the \textit{Cassini}
Composite Infrared Spectrometer (CIRS);
however, the spatial and (in particular) temporal variations in
temperature profiles were minor in these large `beam-footprints'
compared to those seen with the exceptional latitudinal resolution of
\textit{Cassini} \parencite{achterberg_11, vinatier_15,
  coustenis_16}. Here, we seek to use the methodology established in
Paper I to further probe Titan's atmospheric composition and dynamics through the
stratospheric measurements of HCN, HC$_3$N, CH$_3$CN, and C$_3$H$_4$. We present the
first spatially-resolved abundance measurements of these species from
ground-based radio observations, utilizing data
from 2012 to 2015 as Titan transitioned into northern
summer. We discuss the comparisons of these
measurements to those from contemporary \textit{Cassini}/CIRS
observations and photochemical models, and demonstrate the potential
for further studies of spatial and temporal variations in Titan's trace atmospheric
species after the end of the \textit{Cassini} mission using ALMA.

\section{Observations} \label{observations}
We utilize flux calibration data of Titan from the ALMA
science
archive\footnote{https://almascience.nrao.edu/alma-data/archive}, and
follow the procedures in Paper I to reduce and calibrate
datasets. This includes the modification of data reduction scripts
provided by the Joint ALMA observatory to avoid the flagging (removal)
of strong
atmospheric lines from Titan's atmosphere, general imaging procedures, and the
extraction of disk-averaged spectra. The observational parameters for
these data are detailed in Table \ref{tab:obs}. Datasets were chosen
based on spatial resolution and observation date, with preference
given to the highest resolution data observed closest to the data
analyzed in \textcite{thelen_18} used to obtain temperature measurements
in Titan's stratosphere. The list of detected and modeled
transitions for each species is listed in Table \ref{tab:lines}.

We modeled disk-averaged spectra for all datasets listed in Table
\ref{tab:obs}, and spectra from three independent
spatial regions for the nitrile species from either 2012 or 2013, and
for all species in 2014 and 2015. While spectra representing Titan's
northern and southern hemispheres were extracted for C$_3$H$_4$ in 2013, the
signal-to-noise ratio was insufficient to yield meaningful abundance retrievals.
As in Paper I, spatially resolved spectra were extracted from regions where
ALMA `beam-footprints' do not overlap to obtain independent
measurements of three spatial regions. These regions were chosen
to match those in Paper I as closely as possible -- with mean latitudes at or within
3$^\circ$ of $48^\circ$ N,
$21^\circ$ N, and $16^\circ$ S (hereon referred to as `North',
`Center', and `South') -- so that we may ensure corresponding temperature
measurements are appropriate for chemical abundance retrievals. These
regions are held constant with Titan's changing tilt from 2012 to
2015. As we do not expect to see longitudinal changes in chemical
abundance, we extracted spectra representing Titan's low northern latitudes (Center) in some higher resolution data
in 2014 and 2015 from Titan's limb, as opposed to Titan's central disk where emission from
atmospheric species is reduced leading to insufficient fluxes. Integrated flux
(moment 0) maps are
shown for HC$_3$N and CH$_3$CN in Fig. \ref{fig:maps} for 2013, 2014,
and 2015, demonstrating
the spatial variation of these molecules in Titan's stratosphere. 

\section{Spectral Modeling and Retrieval Methodology} \label{results}
Models of Titan's atmospheric structure and the subsequent generation of
synthetic spectra were carried out in a similar fashion to
Paper I. Molecular line data were obtained from the HITRAN 2016\footnote{http://hitran.org/} and CDMS\footnote{https://www.astro.uni-koeln.de/cdms/catalog}
catalogues \parencite{gordon_17, muller_05}. We employed the Non-Linear Optimal Estimator for
Multivariate Spectra Analysis (NEMESIS) radiative transfer code in
line-by-line mode \parencite{irwin_08} to retrieve vertical abundance
profiles from 0--1200 km for each gas species independently. As in
Paper I (and detailed in \cite{teanby_13}), many field-of-view
averaging points (37-44) are required to model disk-averaged data, with
higher concentrations of emission angles on Titan's limb. For
spatially resolved spectra, field-of-view points were weighted to
model emission from each ALMA beam-footprint. 
Spectra were multiplied by a small scaling
factor found by modeling nearby regions of continuum emission to
account for small offsets due to flux calibration or model
inaccuracies. Using accurate troposphere temperatures from
\textit{Cassini} radio science measurements (\cite{schinder_12}, following the
procedures detailed in Paper I) reduced these scaling factors to below $5\%$ of
the continuum flux, within the expected calibration uncertainty level (see ALMA Memo
$\#$594\footnote{https://science.nrao.edu/facilities/alma/aboutALMA/Technology/ALMA$\_$Memo$\_$Series/alma594/memo594.pdf}).
ALMA data and the resulting best fit spectra for each molecule in each year are shown in
Fig. \ref{fig:spec_12}--\ref{fig:spec_15}.

While chemical abundance
contributes to the radiance of molecular line cores in Titan's upper
atmosphere ($\sim$800 km), we only discuss the retrieval results for the upper
troposphere--stratosphere (50--550 km) here, where our previous ALMA retrievals of
temperature are valid (Paper I) and the rotational emission lines are
not subject to non-LTE and thermal broadening
effects \parencite{yelle_91, cordiner_14}; additionally, the few
data points that make up the line centers in these data contribute little to
the $\chi^2$ minimization of data and model discrepancies, and the retrieved
abundance profiles often return to the \textit{a priori} profile
values in the upper atmosphere as in previous studies of ALMA
data using NEMESIS \parencite{serigano_16, molter_16,
  thelen_18}. Contribution functions generated for disk-averaged
spectra of each molecule are shown in Fig. \ref{fig:cf}, showing
significant contribution between $\sim$100--300 km (10--0.1 mbar) for
each molecule, and secondary peaks in the upper atmosphere.

For all gases in this study, multiple \textit{a priori} abundance
profiles were tested to determine uniqueness amongst retrieved
measurements (see example in Section \ref{sec:hc3n}). Similar to the previous temperature retrievals in
Paper I, we have elected to use data besides those
obtained by \textit{Cassini}/CIRS data to generate \textit{a priori}
profiles (i.e. profiles with more simple vertical
structure) where possible, to facilitate the use of ALMA for
future Titan measurements after the end of the \textit{Cassini}
era. These often include previous disk-averaged observations of Titan
in the sub-mm or results from photochemical models. \textit{A priori}
vertical profiles were taken with 100$\%$ errors at all altitudes and
correlation lengths were set to 1.5 scale heights (as in \cite{teanby_07}) for all
species but the HCN isotopologues (set to 3.0, as in \cite{molter_16}) to
provide sufficient vertical smoothing, and to account for
uncertainties due to minor variations in temperature, line broadening
parameters, and ALMA flux calibrations. The specific parameters,
\textit{a priori} profiles, 
and retrieval methodology pertaining to each
gas species are discussed in the following subsections.

To ensure the accurate retrieval of continuous chemical abundance
profiles, we modeled emission lines without any contamination from other species where
possible, and held
atmospheric temperatures constant. Both spatially resolved and disk-averaged temperature profiles from
Paper I were used to model 2012, 2014, and 2015 spectra, providing measurements from
similar latitude regions within $\sim$2 months of the datasets
analyzed here (Table \ref{tab:obs}); temperature
variations in Titan's stratosphere are negligible on these timescales \parencite{flasar_81},
particularly for measurements comprised of multiple latitudes such as
those analyzed here (see Paper I). For 2013 data, we elected
to use more contemporary temperature profiles
obtained during the T89 and T91 \textit{Cassini} flybys (during
February 17 and May 23, respectively) obtained with
the CIRS instrument (courtesy R. Achterberg, private communication;
see also \cite{achterberg_11, achterberg_14}). All temperature
profiles used in this study are are shown in Fig. \ref{fig:temps}.
As emission lines in Titan's atmosphere may be significantly affected
by temperature variations, we tested the discrepancies found in Paper
I between retrieved ALMA and \textit{Cassini}/CIRS
temperature profiles ($\sim$0--5 K for spatial regions) on HCN isotope
lines; we find that variations of temperature
on order 5 K or less resulted in abundance variations \textless
20$\%$, and were well within the retrieval errors.

\subsection{HC$_3$N and C$_3$H$_4$} \label{sec:hc3n}
For both molecules, we assumed a Lorentzian broadening HWHM ($\Gamma$) value = 0.1 cm$^{-1}$
bar$^{-1}$ and temperature dependence ($\alpha$) = 0.75 as in previous studies
\parencite{vinatier_07, cordiner_14, lai_17} and recommended by
HITRAN. The strongest 4-5
C$_3$H$_4$ transitions were modeled each year, as weaker lines did not
significantly contribute to the
retrieved vertical abundance profiles. For the 2015 dataset, 2
interloping C$_2$H$_5$CN lines were modeled among the C$_3$H$_4$
bandhead, except in the Center spectrum, where these lines did not
significantly impact the $\chi^2$ value. 

The initial HC$_3$N abundance profiles used as \textit{a
  priori} inputs came from previous disk-averaged measurements of
Titan in the sub-mm by \textcite{cordiner_14} and
\textcite{marten_02}. A comparison of these profiles to fractional
scale height and continuous abundance retrievals is shown in
Fig. \ref{fig:hc3n}. While adequate fits of disk-averaged lines at low
S/N can be
accomplished using fractional scale height (`gradient') profiles as in \textcite{cordiner_14}, we
find that spectral fits are improved by using continuous abundance
retrievals (Fig. \ref{fig:hc3n}A); for all chemical species, spatial
spectra are fit better by continuous retrievals due to broadened
line wings (compare, e.g., HC$_3$N spectra in
Fig. \ref{fig:spec_14}). Additionally, the gradients present in some
continuous vertical profiles are important for the study of temporal
and dynamical variations. For a variety of \textit{a priori} profiles
(Fig. \ref{fig:hc3n}B) or perturbations thereof, continuous retrievals converged on similar vertical
profiles (Fig. \ref{fig:hc3n}C) for all gases modeled in this study.

Models of C$_3$H$_4$ were initialized using `step models' of abundance,
as in Cordiner et al. (2014; 2015), with a VMR = $1\times10^{-8}$ at 100
km as found by \textcite{nixon_13}, or by using photochemical model
results from \textcite{loison_15}. In 2015,
additional lines of C$_2$H$_5$CN were modeled using the gradient model
from \textcite{cordiner_15},
comparable to that found in other ALMA studies
\parencite{palmer_17, lai_17, teanby_18}. Spatial abundance variations of C$_2$H$_5$CN
are not determined here due to the lines'  proximity to those of
C$_3$H$_4$ and their relatively weak strength. 

\subsection{HCN}
Due to the strong self-absorption present in spectra of HCN from
spatially resolved datasets of Titan and the calibration
uncertainties for species with extensive line wings in ALMA data (as
with CO, detailed in Paper I), we chose to model the HCN isotopologues
H$^{13}$CN and HC$^{15}$N as proxies for HCN
abundance. \textcite{molter_16} showed a retrieved vertical abundance
profile of HCN could be scaled to fit
lines of isotopologues and used to determine isotope ratios for disk-averaged
spectra. Here, we
reversed this process by fitting H$^{13}$CN and HC$^{15}$N lines using
the HCN profile found by \textcite{molter_16}, and applied a
constant scaling factor ($^{12}$C/$^{13}$C =
89.8, $^{14}$N/$^{15}$N = 72.2) to determine the HCN
abundances. As in that study, we model H$^{13}$CN and HC$^{15}$N lines
with $\Gamma$ = 0.13, $\alpha$ = 0.75. We set condensation to begin at
altitudes below $\sim$ 80 km as in \textcite{marten_02}, which was
derived from the vapor saturation law in
\textcite{lellouch_94} and is consistent with
the calculations based on \textit{Cassini/Huygens}
observations in Titan's lower stratosphere by \textcite{lavvas_11b}; below
this altitude, the abundance profile no
longer affects the line shape. The HCN isotope lines lie on the wings of the CO
(J=3--2) and CO (J=6--5) transitions, so those lines were included in the
model using the parameters given in Paper I.

While the C and N ratios found for Titan through HCN
measurements have a range of values (see \cite{molter_16} and
references therein), we find that applying these ratios to line data
and scaling the retrieved profiles to convert to HCN
abundances have vanishingly small effects for the range of published
isotope ratios. For the 2014 measurements, we were able to model both
species in disk-averaged and spatially resolved spectra, and found the
(scaled) retrieved profiles were in good agreement (see Section \ref{sec:disc}).

\subsection{CH$_3$CN}
For computational efficiency and to preserve the native resolution of
ALMA data (i.e. without additional channel averaging to meet the array limitations of
NEMESIS), we only retrieved abundance profiles using the
strongest CH$_3$CN lines in each band. Studies of CH$_3$CN in Titan's atmosphere have been limited due to its
lack of observable transitions in the IR accessible by
\textit{Cassini}. Therefore we tested three different \textit{a
  priori} profiles and variations of those by an order of magnitude in
each direction: the combination of disk-averaged measurements from
\textcite{marten_02} to 500 km, and the model results from
\textcite{loison_15} up to 1200 km; the model by
\textcite{dobrijevic_18}, which tests a new nitrogen isotope
fractionation scheme from \cite{loison_15}; a test gradient profile (Profile 1 from
Fig. \ref{fig:hc3n}). \textit{A priori} profiles and retrieval results are
shown in Fig. \ref{fig:ch3cn}A and B, respectively, for the 2014
disk-averaged spectrum of CH$_3$CN (Fig. \ref{fig:spec_14}).
All retrievals shown in Fig. \ref{fig:ch3cn}B provided an adequate fit
to the data. We find that the retrievals converge
around 150 km ($\sim2$ mbar) and above, where ALMA is sensitive to CH$_3$CN emission
(Fig. \ref{fig:cf}E). 

We adopted the N$_2$-broadening
parameters detailed by \textcite{dudaryonok_15b}, where available. As
these differ from the parameters used by \textcite{marten_02}, we ran
a large number of forward models of the 2014 (J=16--15)
transitions to test the effect of the Lorentzian broadening and
temperature dependence coefficients. Though we obtained some variation
in $\chi^2$ values for the parameter space [$\Gamma$=0.1--0.16,
$\alpha$=0.5--0.8] for CH$_3$CN forward models, the effects of these
parameters on retrieved abundances were small and well within the
retrieval errors for a model using the \textcite{dudaryonok_15b}
parameters.

\section{Results and Discussion} \label{sec:disc}
In Fig. \ref{fig:abda}, we present the mean disk-averaged results for
each molecule; the average
of the scaled HC$^{15}$N and H$^{13}$CN profiles is used to represent
HCN here. As with the temperature profiles found in Paper I,
abundance retrievals from disk-averaged measurements do not show
significant variation from year to year, and all fall within the
retrieval errors of the mean profile. In
Fig. \ref{fig:ab12}--\ref{fig:ab15}, we present the retrieved
abundance profiles from spatially resolved spectra in
Fig. \ref{fig:spec_12}--\ref{fig:spec_15}. HCN profiles from 2012 and
HC$_3$N, CH$_3$CN profiles from 2013 are shown together in
Fig. \ref{fig:ab12}. Scaled HCN profiles from both H$^{13}$CN and
HC$^{15}$N retrievals from 2014 are shown in Fig. \ref{fig:ab14}. With
the exception of a small portion of the Center retrievals between
0.1--1 mbar, and \textgreater10 mbar (where the HCN isotopologues quickly
lose sensitivity -- Fig. \ref{fig:cf}A,B), these profiles all
agree and display similar vertical variations (e.g. a slight
inversion in north and south retrievals at pressures \textless0.1 mbar).

\subsection{Comparison to Previous Studies}
%Disk-averaged
In Fig. \ref{fig:da_comp} we compare the mean disk-averaged profiles
(Fig. \ref{fig:abda}) to those from previous disk-averaged sub-mm
measurements of Titan and photochemical model results.

%HCN
The disk-averaged HCN profile (a mean of both H$^{13}$CN and
HC$^{15}$N profiles for all years) agrees well with previous sub-mm
observations by \textcite{molter_16} with ALMA throughout the
atmosphere, and with those of
\textcite{marten_02} and \textcite{gurwell_04} in the lower
atmosphere. Our profile is also comparable to the photochemical models
of \textcite{krasnopolsky_14} and \textcite{dobrijevic_18} in
the stratosphere and above.

%HC3N
Our mean retrieved HC$_3$N profile shows a highly variable slope --
particularly the lower atmosphere enhancement near 1 mbar -- compared to both previous
sub-mm observations \parencite{marten_02, cordiner_14} and
photochemical models \parencite{dobrijevic_18}, though the abundance
at all altitudes is significantly
less than predicted by \textcite{krasnopolsky_14}. We find
stratospheric abundances closest to the fractional scale
height model adopted by \textcite{cordiner_14} and the models of
\textcite{dobrijevic_18}. These differences may be explained by the use of
continuous abundance retrievals for disk-averaged measurements, which
tended towards a mean profile of the three spatial regions; HC$_3$N
shows significant enhancement between 100--200 km in the higher
northern and low southern latitudes
(Fig. \ref{fig:ab12}--\ref{fig:ab15}), which may be reflected in the
disk-averaged measurements.

%C3H4
The mean C$_3$H$_4$ profile is consistent with previous CIRS
measurements \parencite{nixon_13}, contemporary ALMA
observations \parencite{teanby_18}, and photochemical
models \parencite{krasnopolsky_14, loison_15} below 400 km, where ALMA
is most sensitive to C$_3$H$_4$ emission (Fig. \ref{fig:cf}D).

%CH3CN
Above 200 km, we find that our CH$_3$CN profile is consistent with previous sub-mm observations by
\textcite{marten_02} and the photochemical model of
\textcite{dobrijevic_18}, though generally less than that of
\textcite{krasnopolsky_14}. We find CH$_3$CN to be a factor of $\sim5$
less than the upper limit found by \textcite{nixon_10} at 25$^\circ$S
in \textit{Cassini}/CIRS measurements at 0.27 mbar. Near 1 mbar, our retrieval results and the
other profiles shown in Fig. \ref{fig:da_comp} diverge, which may be
indicative of another loss mechanism for CH$_3$CN in Titan's lower
stratosphere that has not been accounted for by photochemical
models. At higher pressures (particularly \textgreater10 mbar,
or \textless100 km) our retrievals adhere more
strongly to the input vertical profiles (Fig. \ref{fig:ch3cn}), inhibiting us from accurately
determining the nature of CH$_3$CN's lower atmosphere gradient.

%Cassini Comparisons
Our retrievals are compared to contemporary \textit{Cassini}/CIRS
limb \parencite{vinatier_15} and nadir \parencite{coustenis_16}
measurements in Fig. \ref{fig:c_comp}.
We measure lower abundances than those found by
\textcite{coustenis_16} from 50$^\circ$N and S
nadir observations at the peak of their contribution functions at 7
(HCN) or 10 (HC$_3$N and C$_3$H$_4$) mbar; however, these nadir
measurements are assuming constant vertical profiles above
condensation altitudes, where our continuous retrievals often manifest
as steep gradients in the lower atmosphere. Our results agree better at altitudes above those sounded
by 2012 and 2013 CIRS nadir observations,
particularly at the altitudes of the secondary peaks in the contribution functions of
HCN and HC$_3$N near 0.1--0.5 mbar seen in 2014. 

%HCN
We find our 2012 HCN profiles (derived from HC$^{15}$N) to be
comparable to the CIRS limb measurements by \textcite{vinatier_15} in all
regions, with the exception of the south at pressures \textless0.01 mbar; here, the
large latitudinal average of our ALMA beam-footprint measurements may
be less directly comparable to CIRS, which is more sensitive to variations in
the upper atmosphere. Due to the
\textit{Cassini} orbiter's high
latitude resolution and preferable viewing geometry, abundance
enhancements as a result of subsidence onto the south pole, or the increased
formation/decreased destruction of these molecules in southern winter,
are more readily apparent. Further, sub-mm
observations lose sensitivity in the upper atmosphere ($\textgreater800$
km) for all
molecules observed here (Fig. \ref{fig:cf}). For example, while the
northern CIRS profile falls within our retrieval errors, we do not
observe the same vertical structure in the upper atmosphere showing a
depletion of HCN at pressures \textless0.02 mbar, as our retrieved profile tends
to adhere more strongly to the \textit{a priori} values in the upper atmosphere.

%HC3N
Similar discrepancies are
observed for HC$_3$N, where we find a more shallow gradient at low northern (Center) and southern (South)
latitudes at high altitudes compared to the 2012 CIRS retrievals. We might expect that the
relatively large ALMA beams may more easily obfuscate spatial variations in
shorter lived trace species, such as
HC$_3$N and C$_3$H$_4$, which are more susceptible to short term
change in atmospheric circulation or increased
production as the moon transitions into southern winter. 
As HC$_3$N seems to be a good tracer of atmosphere dynamics in Titan's
stratosphere (Fig. \ref{fig:maps}), continued ALMA monitoring of
this molecule, particularly with higher spatial resolution, may
help elucidate changes in shorter lived nitriles and circulation in
the stratosphere. 

%C3H4
Due to the lack of earlier spatially resolved C$_3$H$_4$ observations,
we compare our 2014 retrievals to those of
\textcite{vinatier_15} from 2012. Unlike in HCN and HC$_3$N, the ALMA-
and CIRS-derived
southern profiles here are in good agreement, as are the measurements
from low northern latitudes (Center). The lower altitude enhancement at mid-northern latitudes
(North) rises by $\sim$30 km (from 170--200 km), and
increases in magnitude by a factor of 2.8. This may not be
unreasonable, as \textcite{vinatier_15} and \textcite{coustenis_16}
both observe a general increase in C$_3$H$_4$ abundance at mid-northern
latitudes into northern spring, and the now reversed pole-to-pole
circulation cell may shift a lower atmosphere reservoir of C$_3$H$_4$ to
higher altitudes. As with the other gases, the abundance measurements
derived from ALMA observations comprise multiple latitude decades,
making direct comparisons to CIRS limb observations difficult;
however, we find that our results are generally
compatible with those from \textit{Cassini}, previous ground-based
observations, and photochemical model results, particularly near 1--10
mbar, where our retrievals are most sensitive.

\subsection{Spatial and Temporal Variations}
\begin{figure}
\subfigure{\includegraphics[width=6cm]{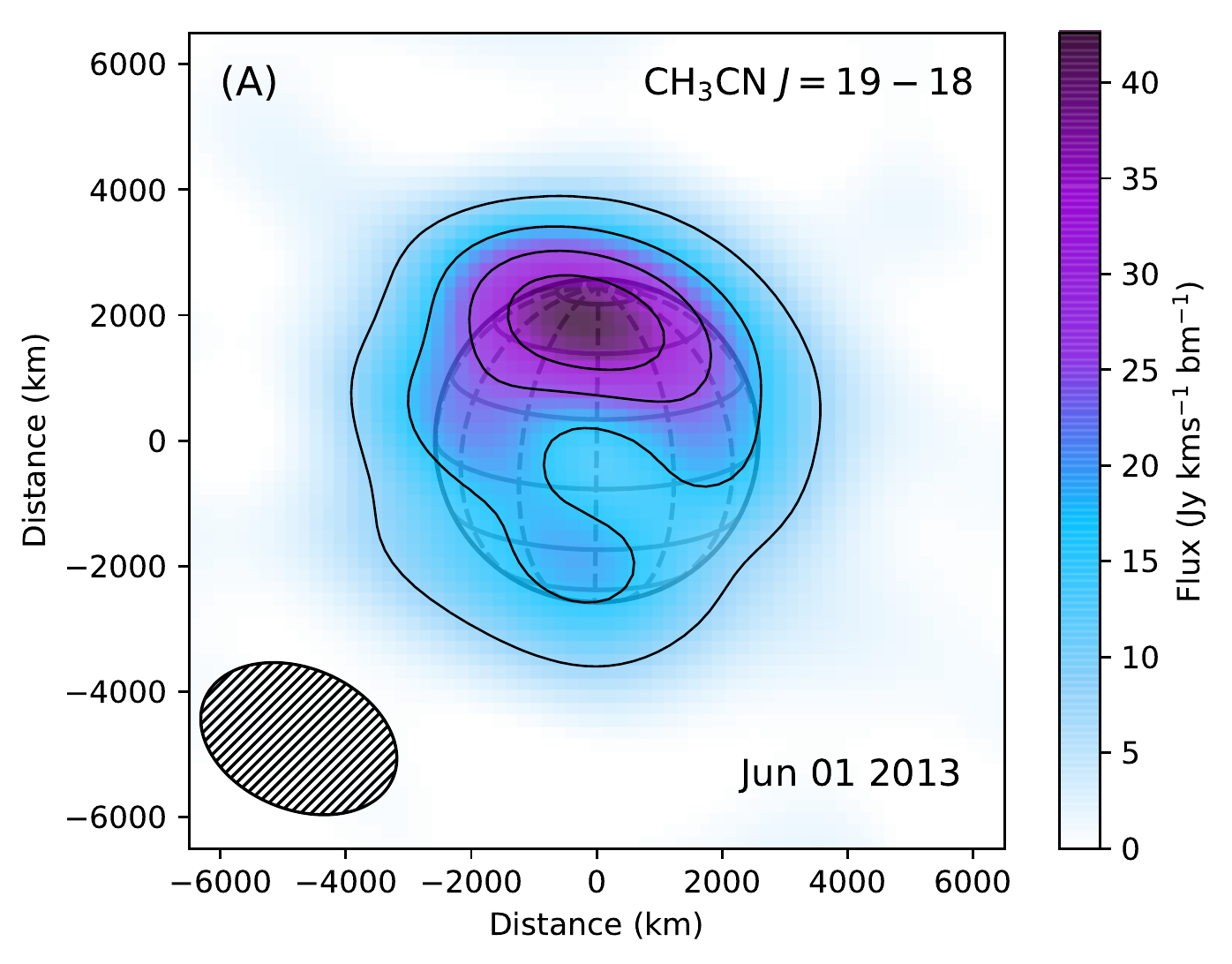}}
\subfigure{\includegraphics[width=6cm]{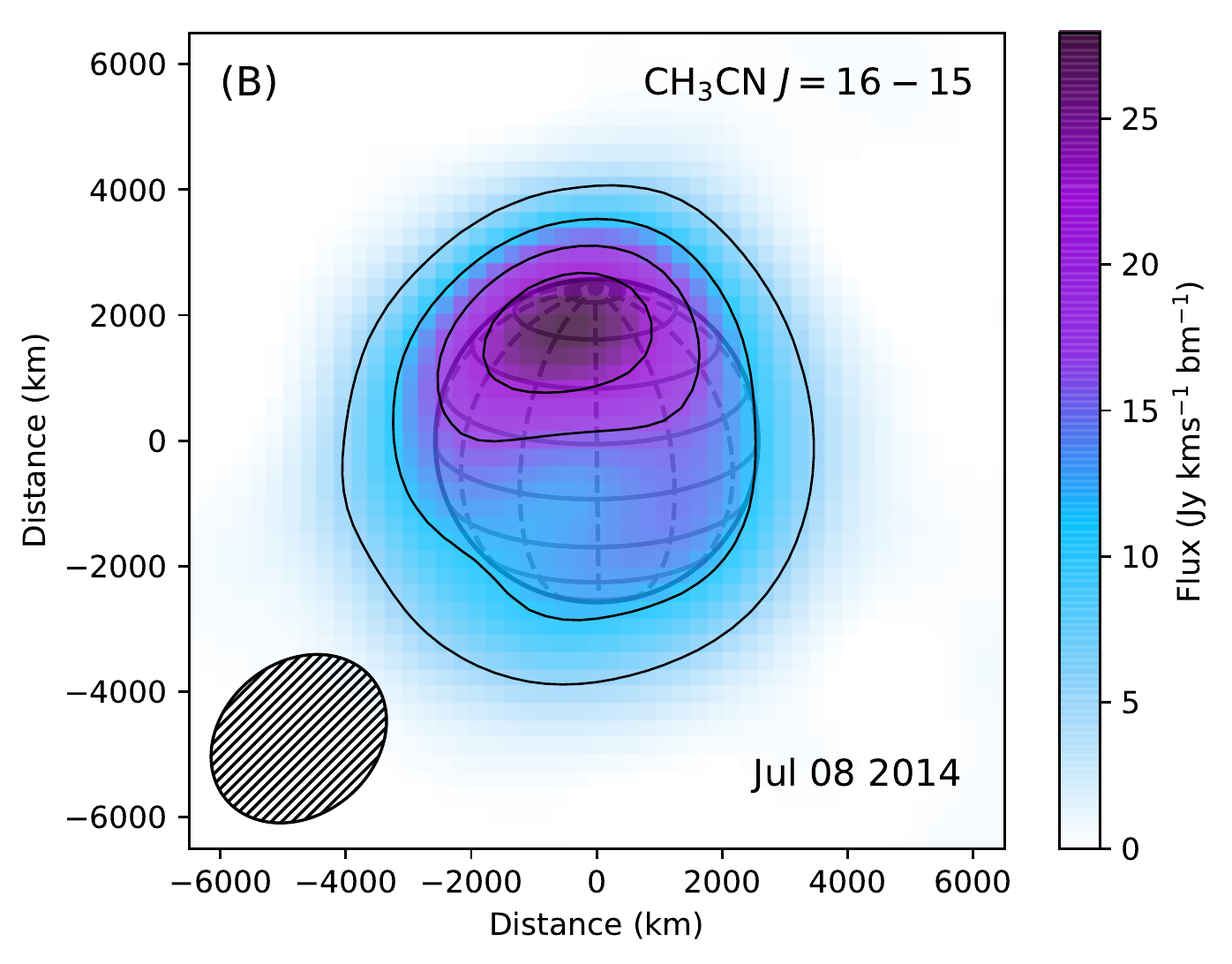}}
\subfigure{\includegraphics[width=6cm]{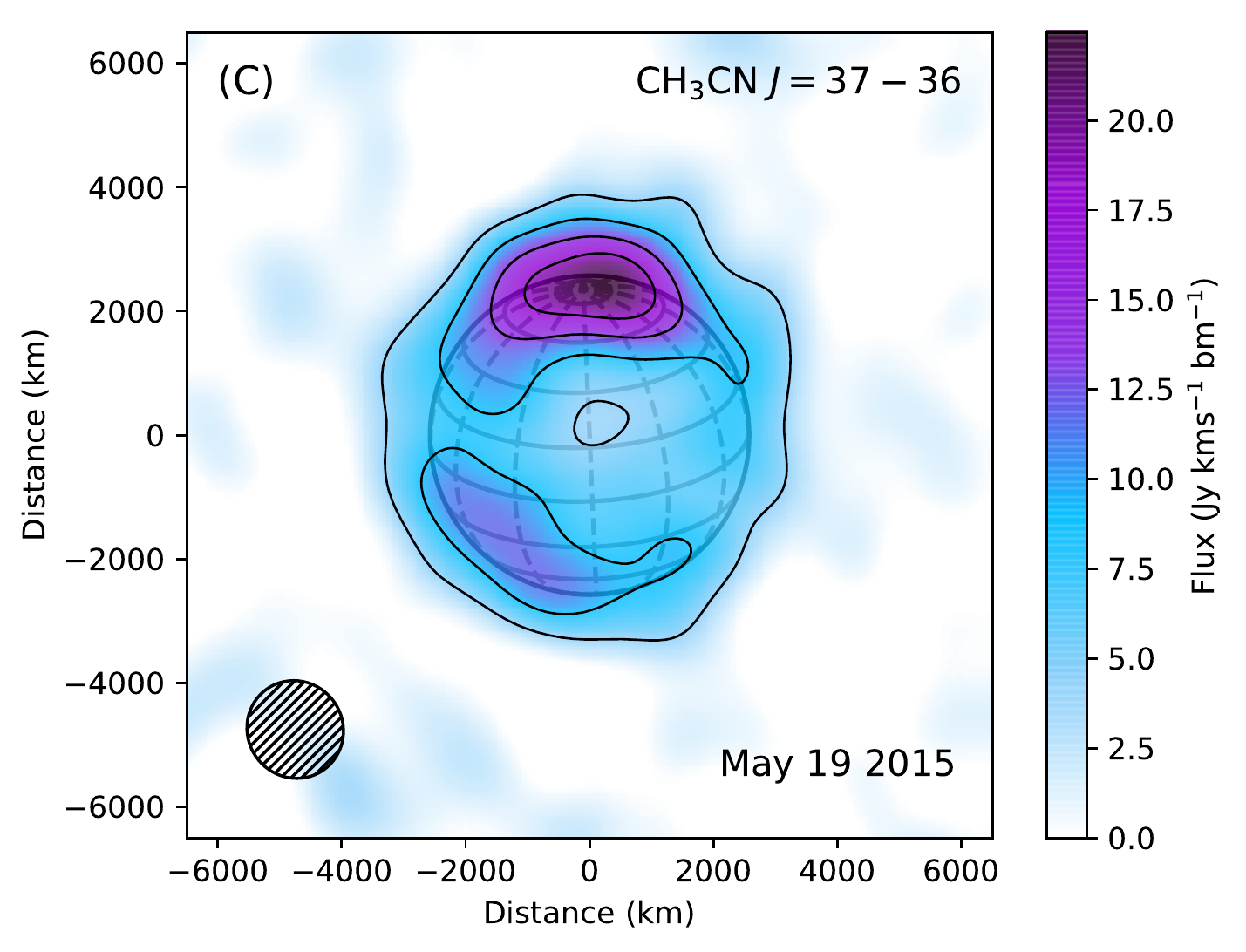}}
\subfigure{\includegraphics[width=6cm]{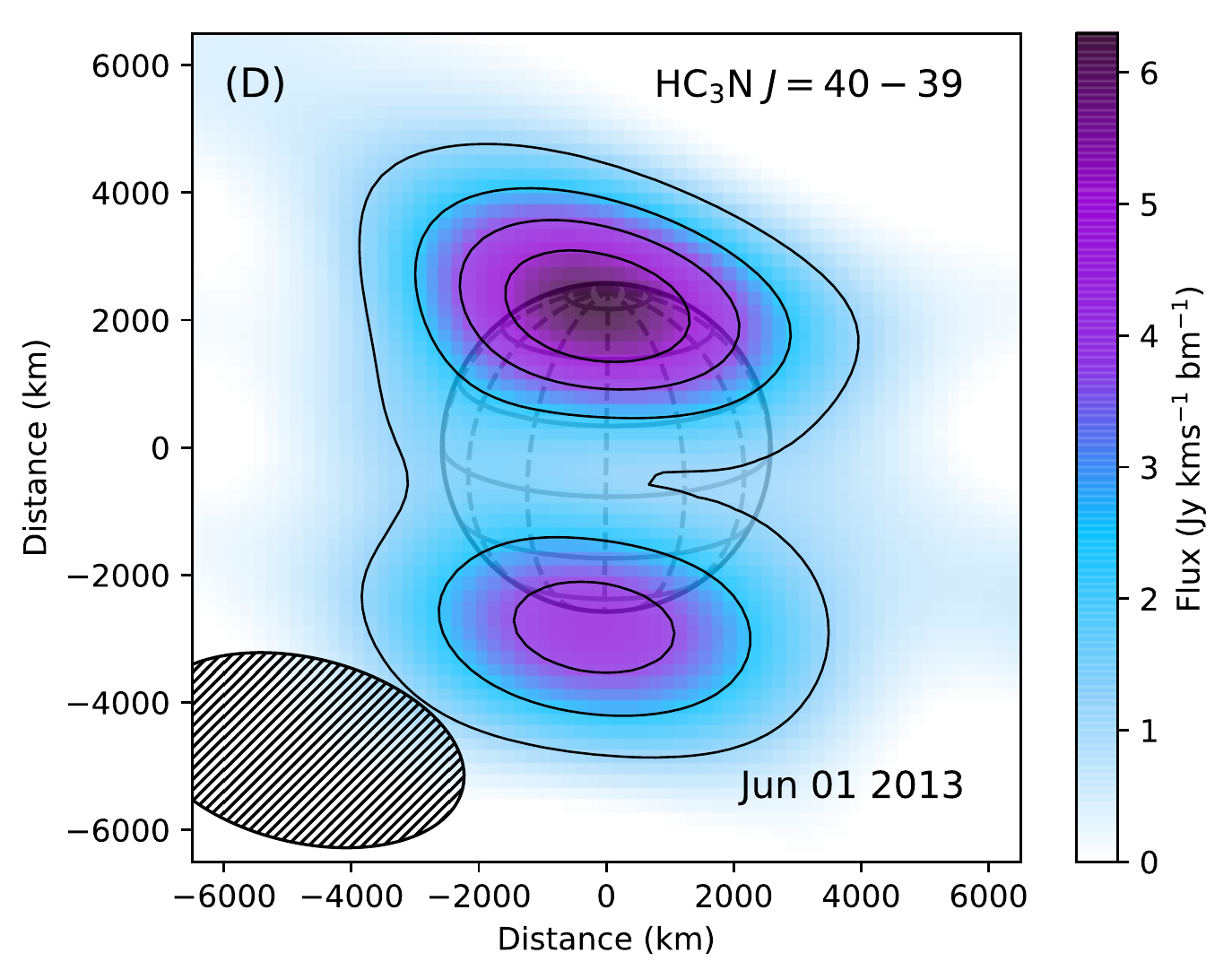}}
\subfigure{\includegraphics[width=6cm]{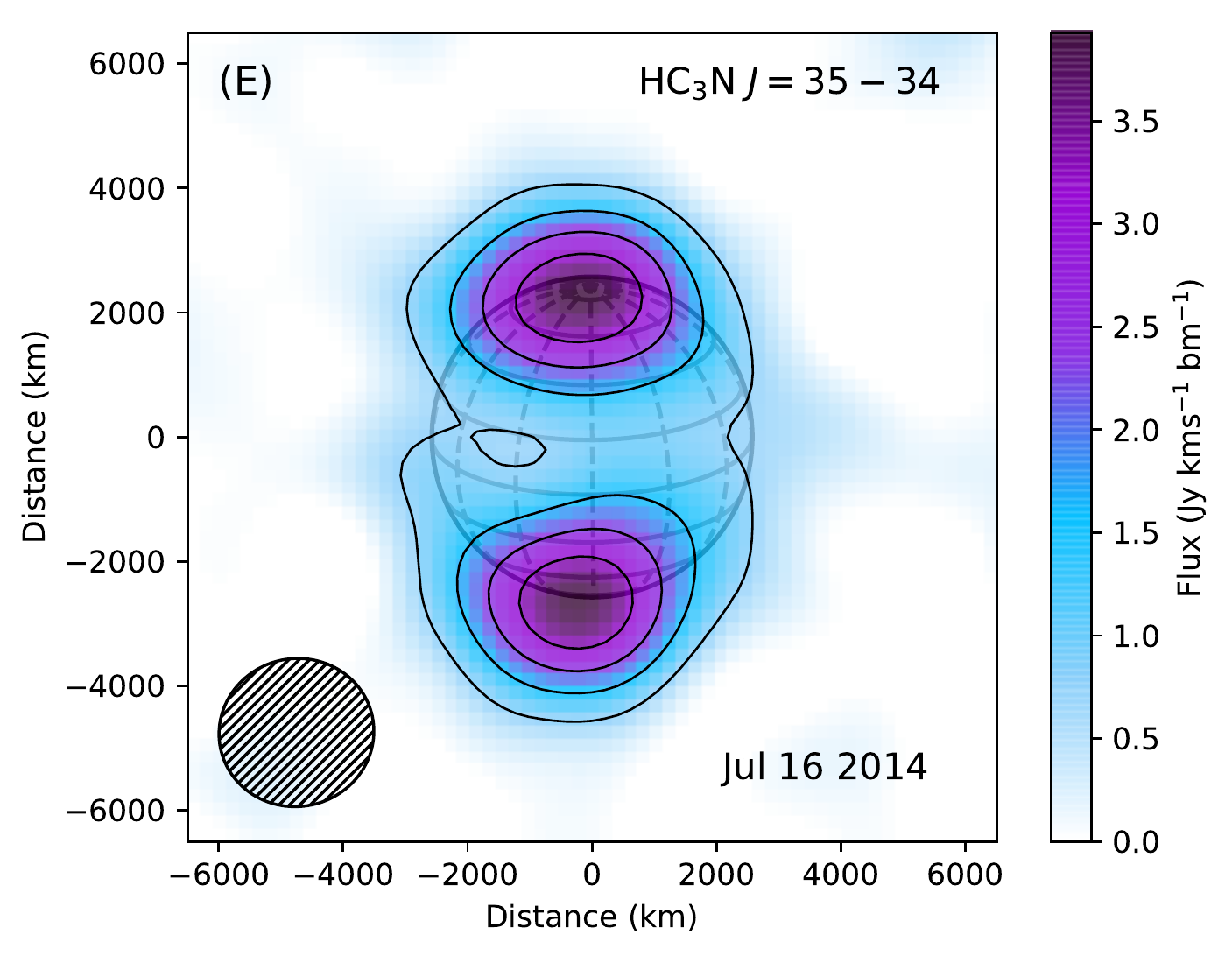}}
\subfigure{\includegraphics[width=6cm]{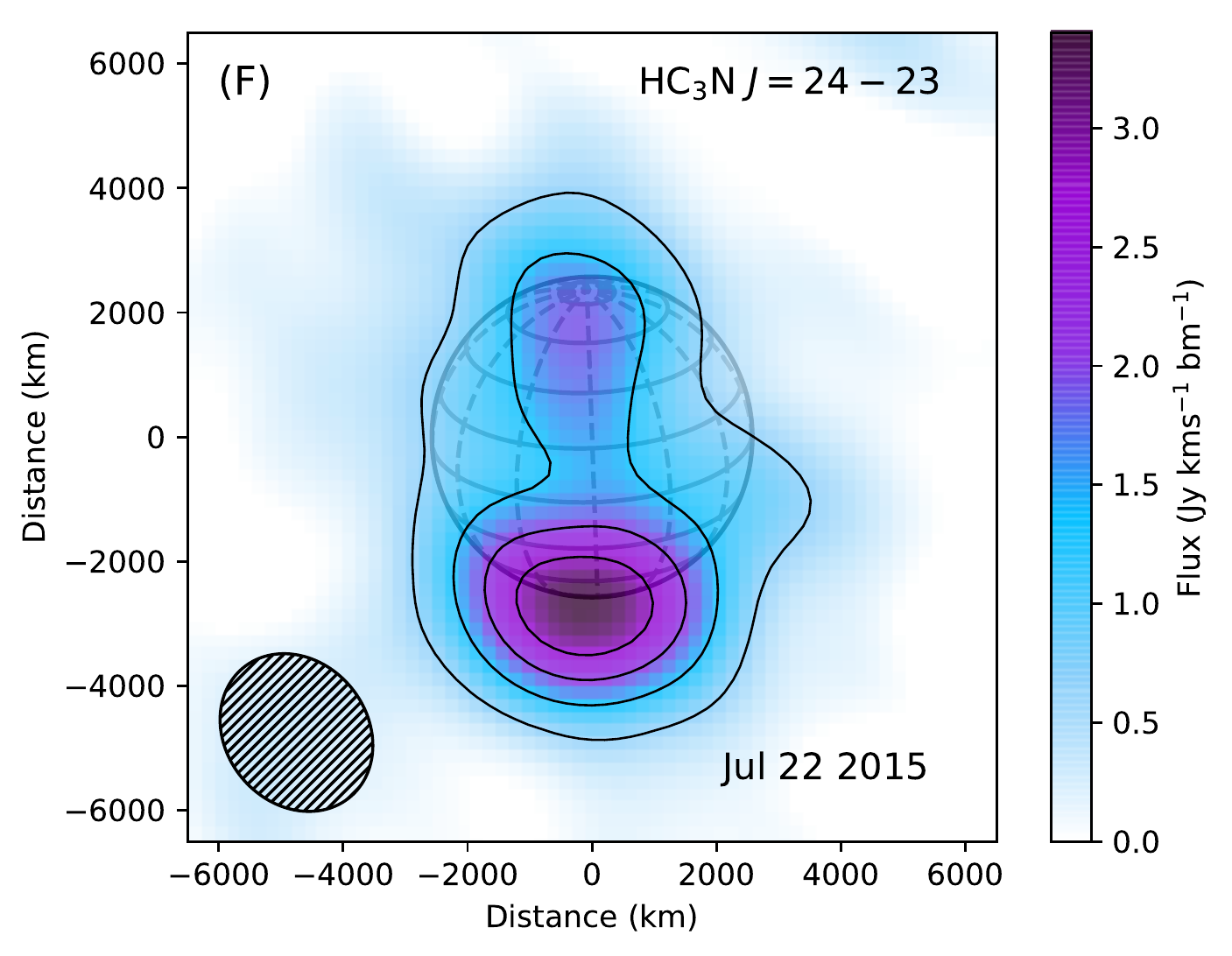}}
\caption{Maps of integrated flux for CH$_3$CN (A--C) and HC$_3$N
  (D--F) lines from 2013, 2014, and 2015. Contours are in intervals of
$F_m/5$ (where $F_m$ is the maximum flux of each image cube). The solid
gray circle denotes Titan's surface, with lines for latitude (solid)
and longitude (dashed) in 22.5$^\circ$ and 30$^\circ$ increments,
respectively. Hashed ellipses represent the FWHM of the Gaussian restoring beam
for each ALMA observation (see Table \ref{tab:obs}).}
\label{fig:maps}
\end{figure}

\begin{figure}
\begin{center}
\subfigure{\includegraphics[scale=1.]{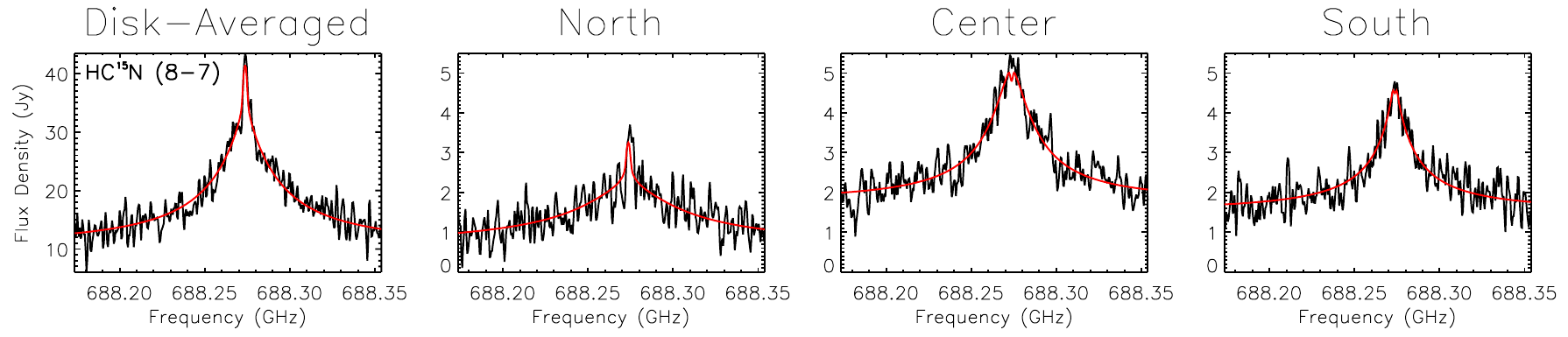}}
\caption{Disk-averaged (first panel), northern (second), central
(third), and southern (fourth) ALMA spectra (black) and synthetic best fit spectra (red) are
shown for 2012 HC$^{15}$N emission lines. Spatial spectra are shown on
the same y-axis scale to illustrate differences in flux density
between various latitudinal regions.}
\label{fig:spec_12}
\end{center}
\end{figure}

\begin{figure}
\begin{center}
\subfigure{\includegraphics[scale=0.93]{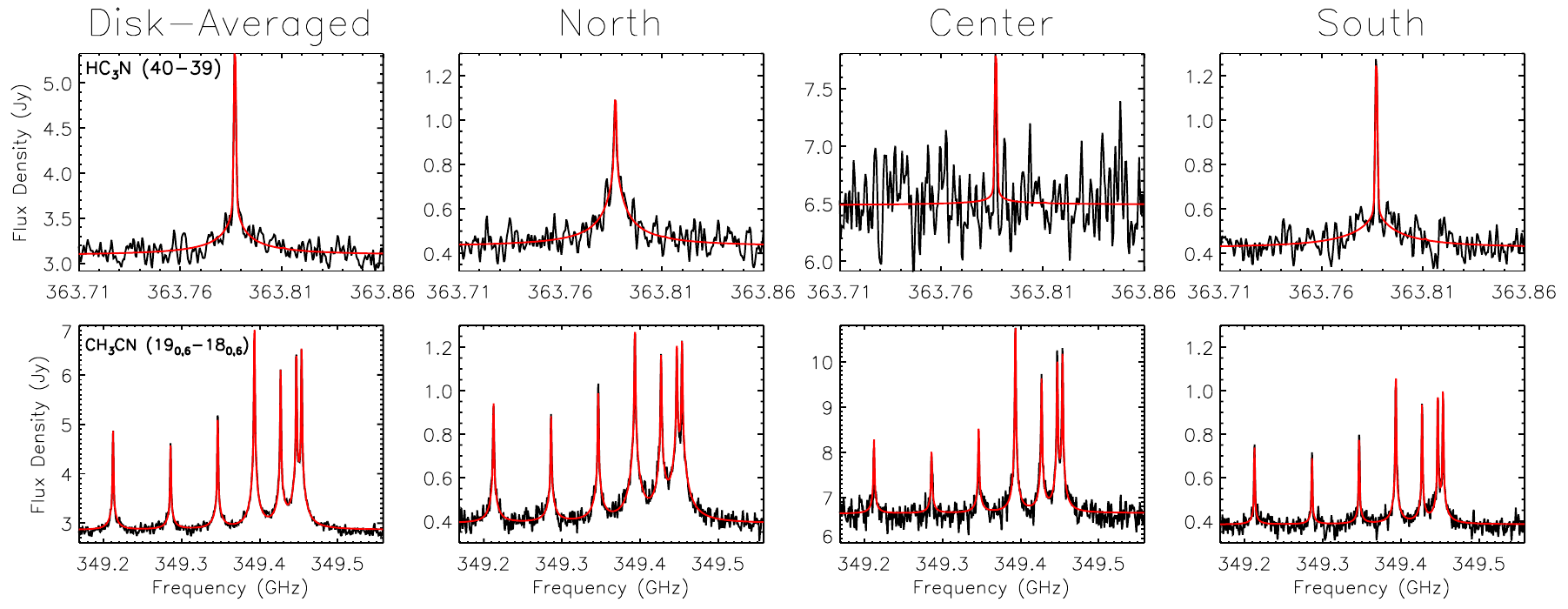}}
\caption{2013 spectra and best fit models are shown for HC$_3$N (top
  row), and CH$_3$CN (bottom) as in
  Fig. \ref{fig:spec_12}. Center spectra are on a different y-axis
  scale than the other spatial spectra due to the large variations in flux
  density as a function of beam size.}
\label{fig:spec_13}
\end{center}
\end{figure}

\begin{figure}
\begin{center}
\subfigure{\includegraphics[scale=0.93]{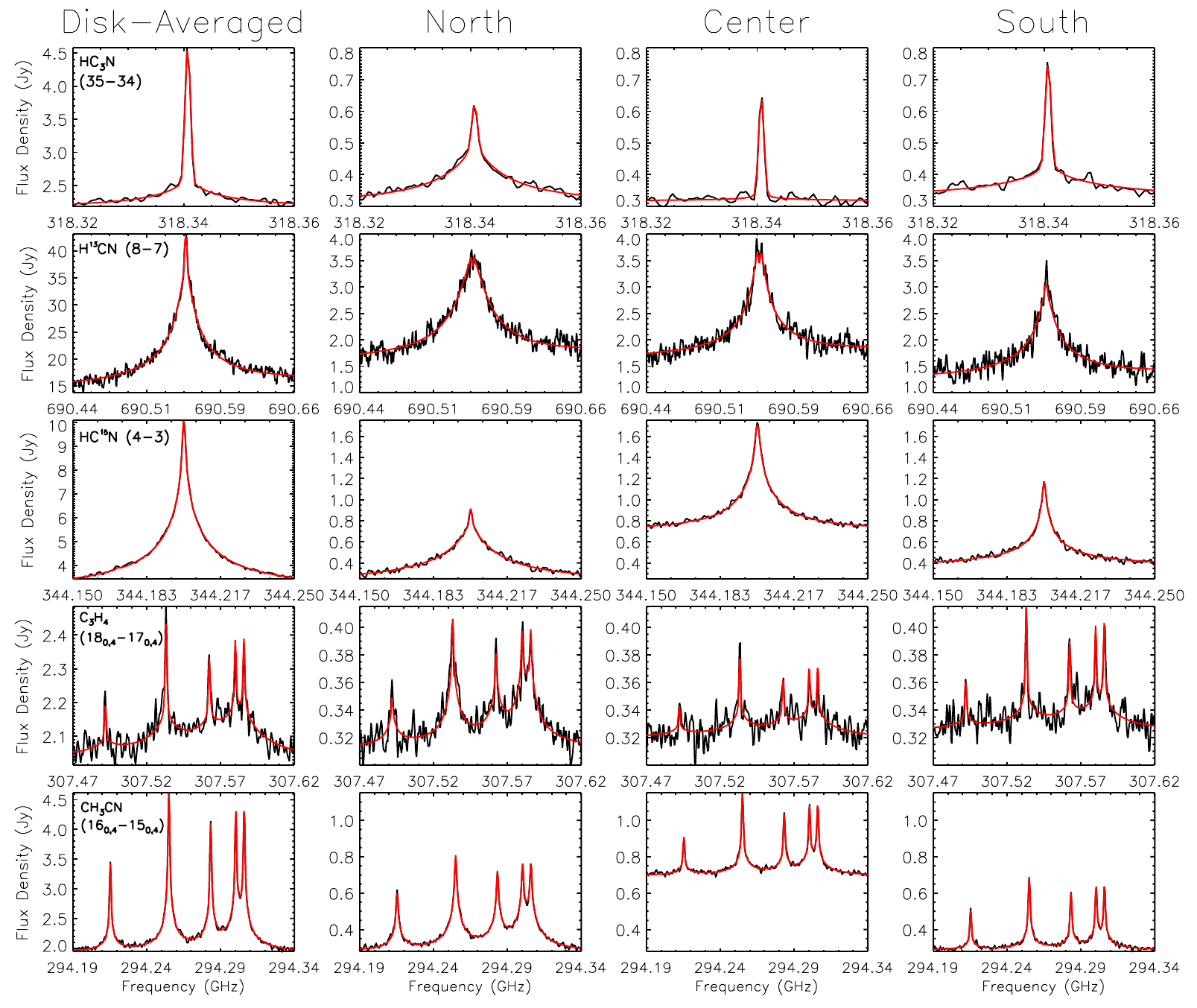}}
\caption{2014 spectra and best fit models are shown for HC$_3$N (first
  row), H$^{13}$CN (second), HC$^{15}$N (third), C$_3$H$_4$ (fourth), and CH$_3$CN (fifth)
  as in Fig. \ref{fig:spec_12}.}
\label{fig:spec_14}
\end{center}
\end{figure}

\begin{figure}
\begin{center}
\subfigure{\includegraphics[scale=1]{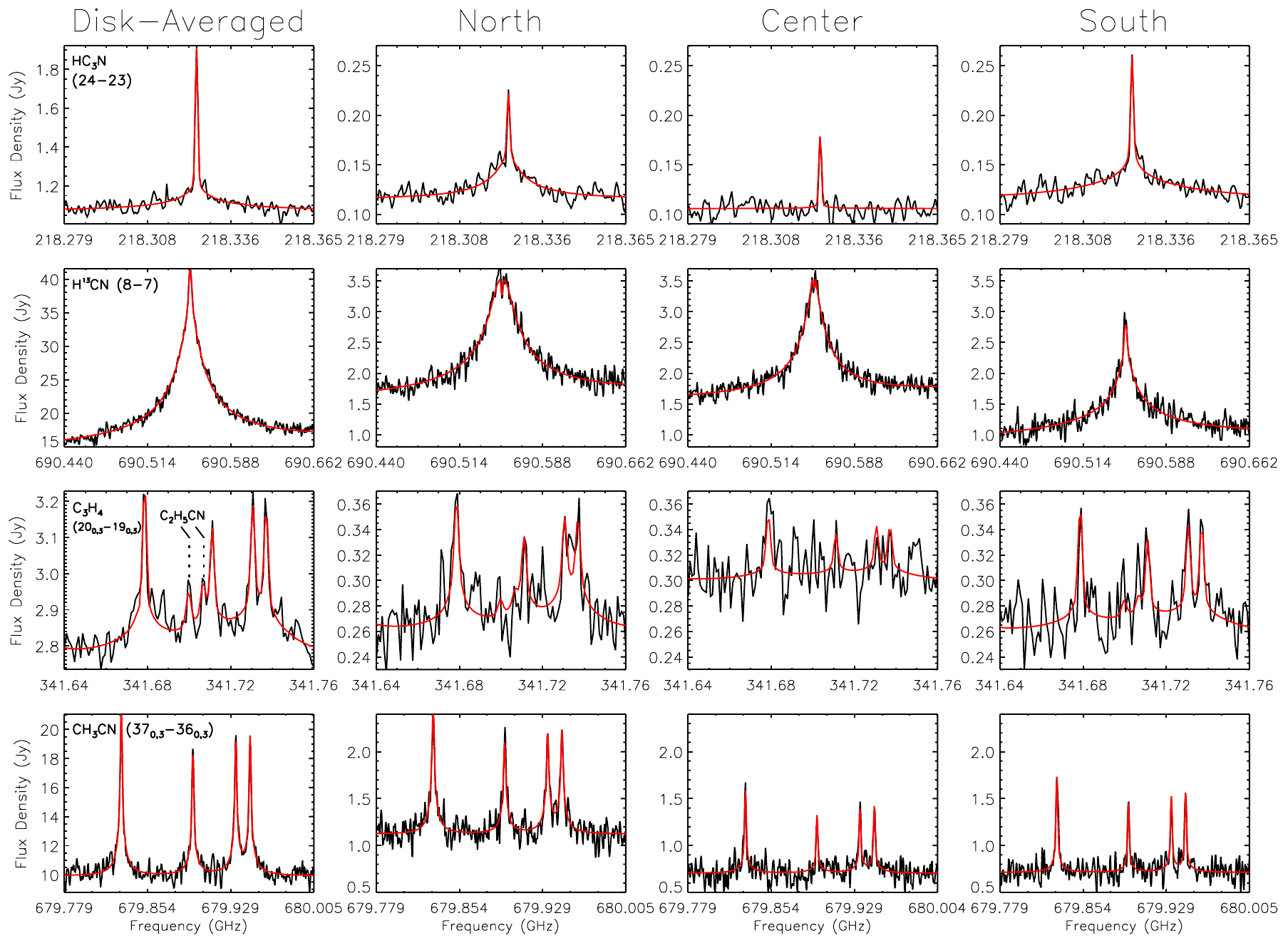}}
\caption{2015 spectra and best fit models are shown for HC$_3$N (first
  row), H$^{13}$CN (second), C$_3$H$_4$ (third), and CH$_3$CN (fourth)
  as in Fig. \ref{fig:spec_12}. Interloping C$_2$H$_5$CN lines in the
  C$_3$H$_4$ spectra are shown with dotted lines in the disk-averaged panel.}
\label{fig:spec_15}
\end{center}
\end{figure}

\begin{figure}
\subfigure{\includegraphics[scale=0.5]{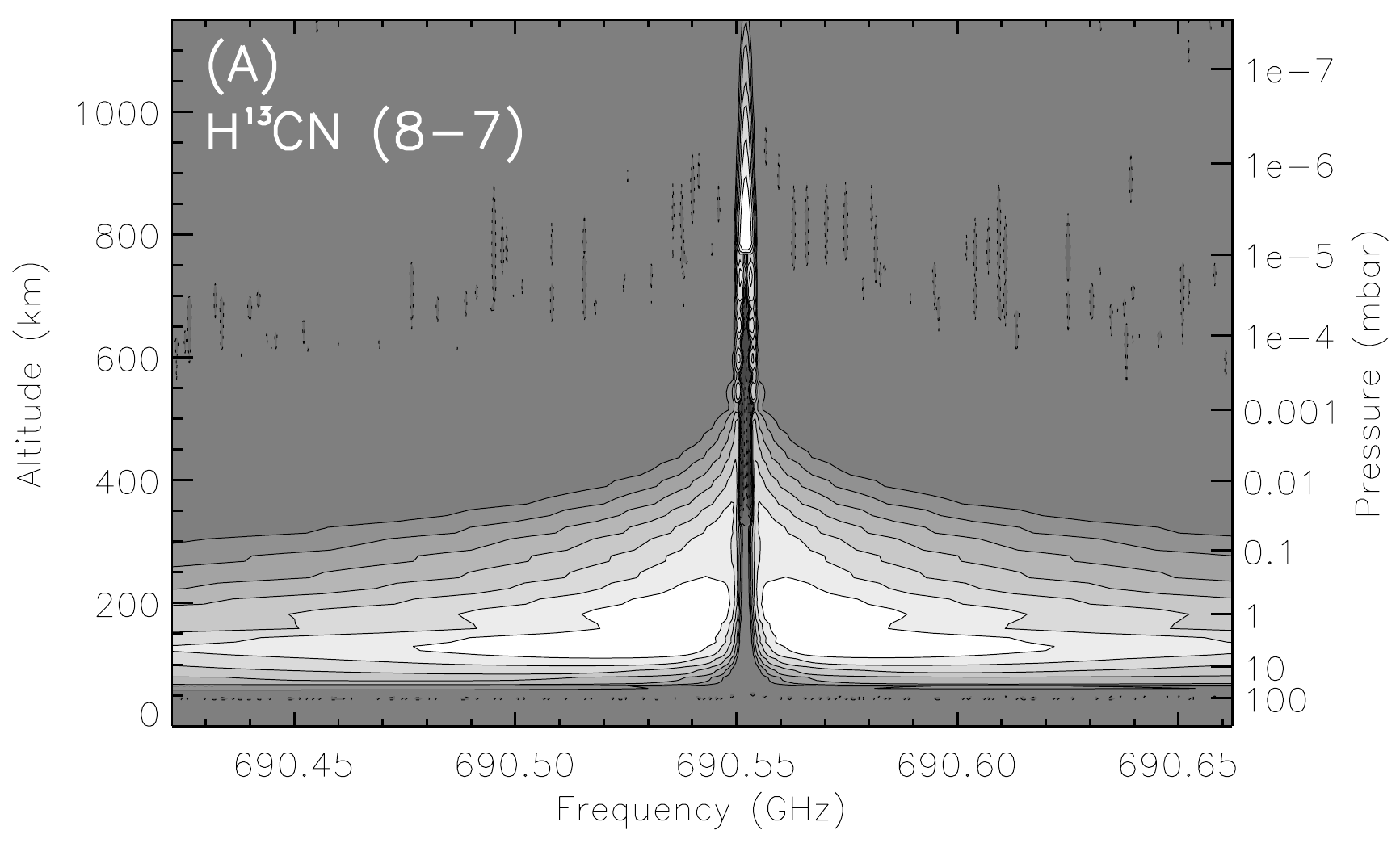}}
\subfigure{\includegraphics[scale=0.5]{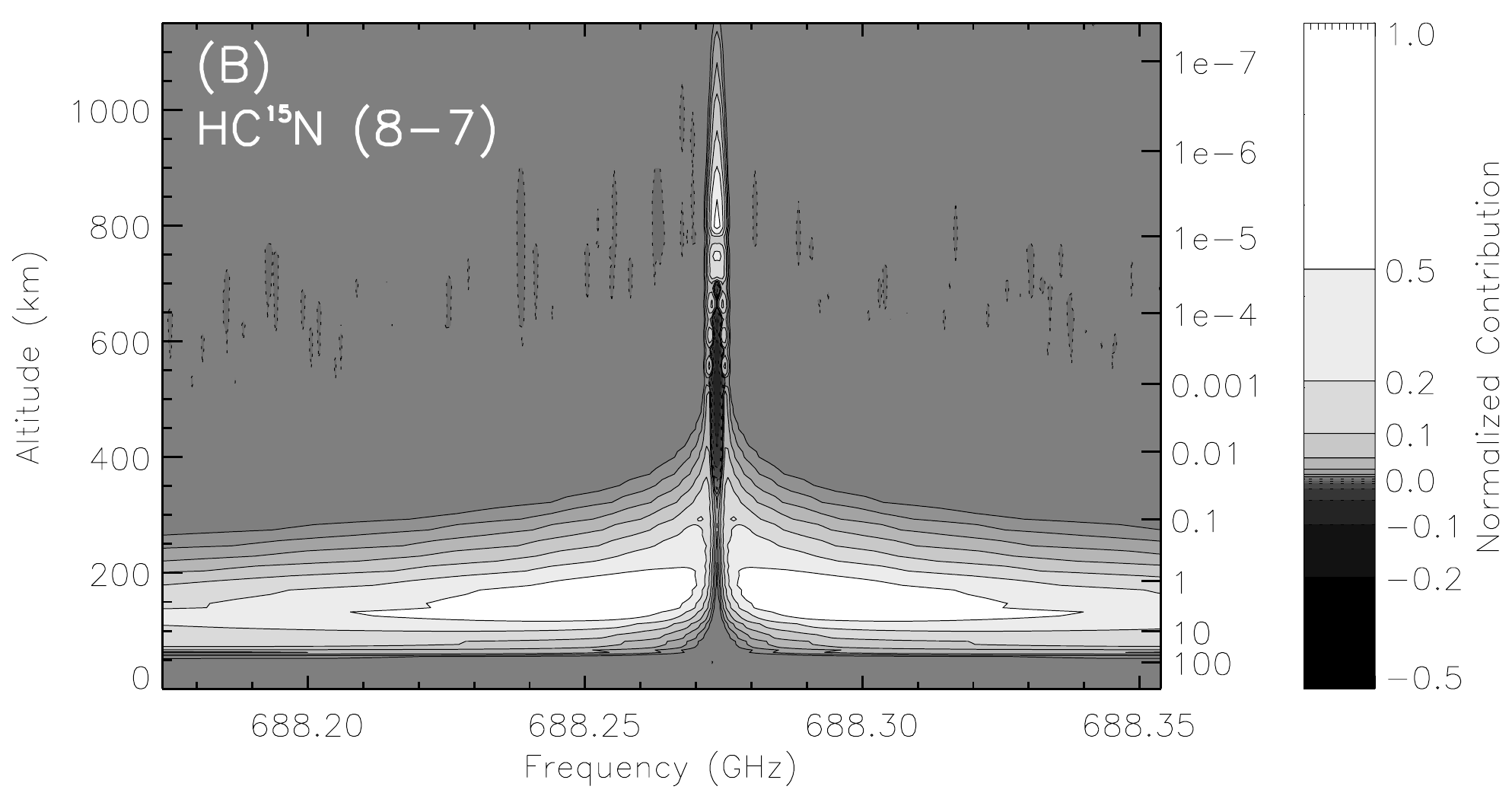}}
\subfigure{\includegraphics[scale=0.5]{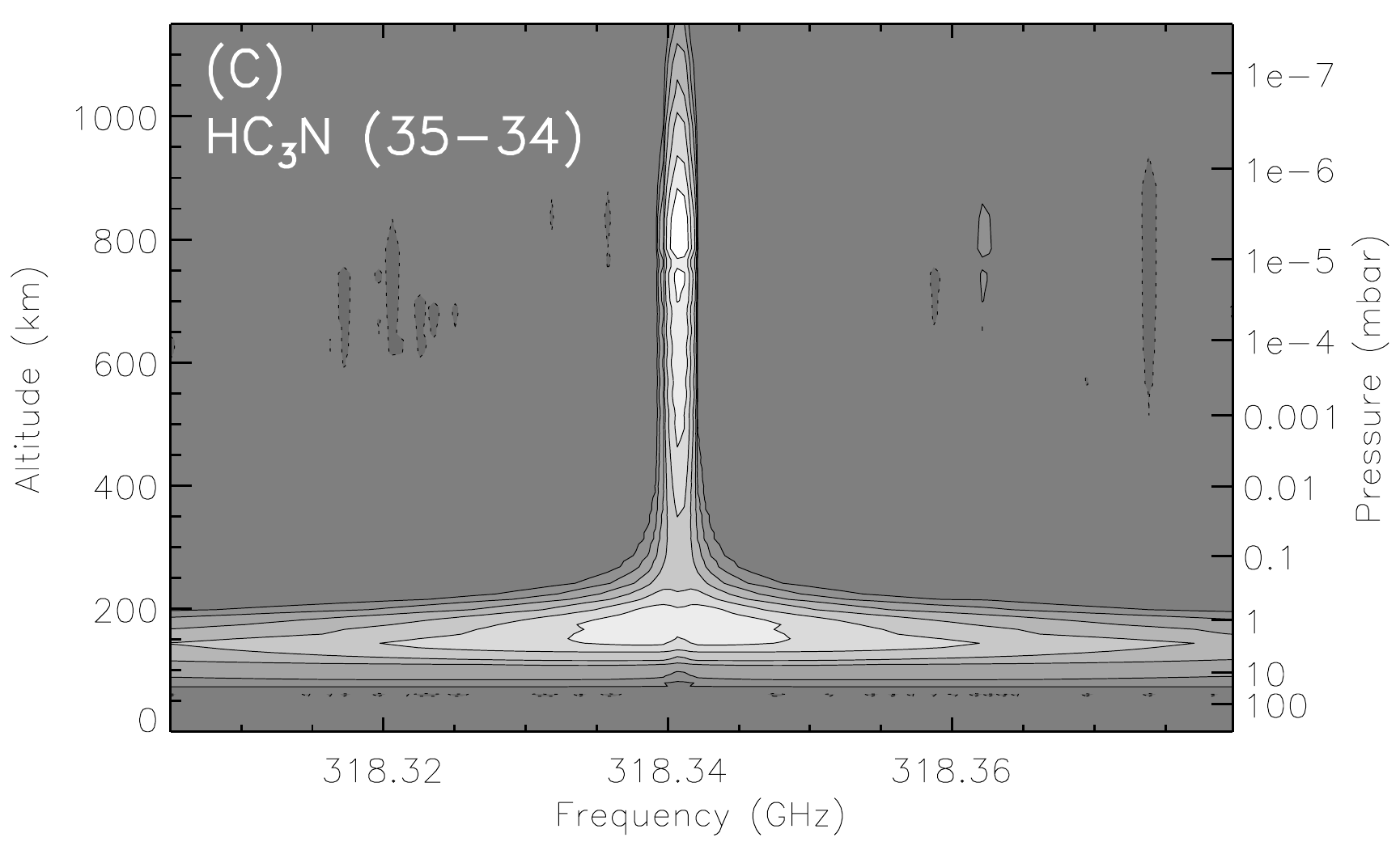}}
\subfigure{\includegraphics[scale=0.5]{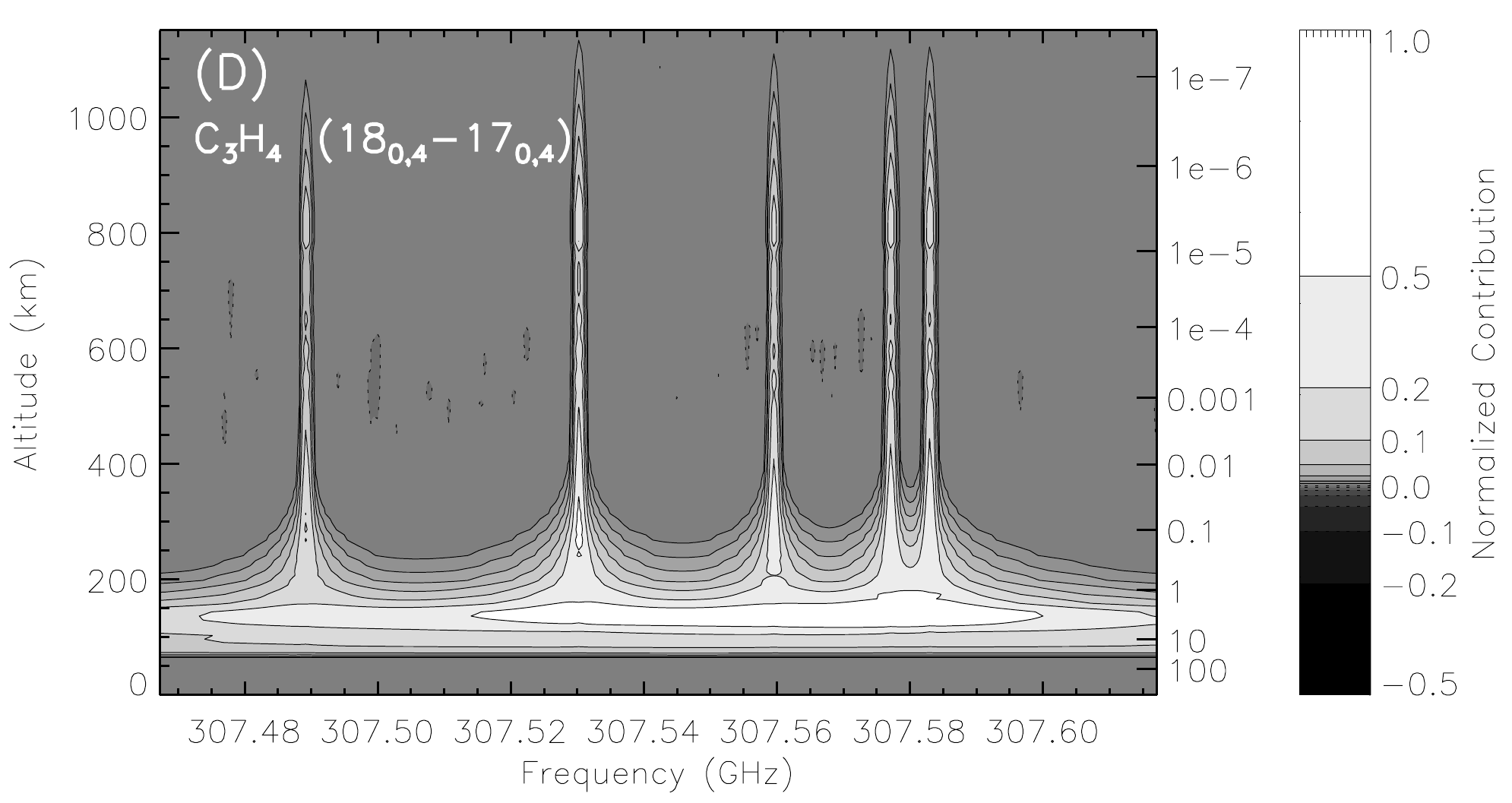}}
\begin{center}
\subfigure{\includegraphics[scale=0.5]{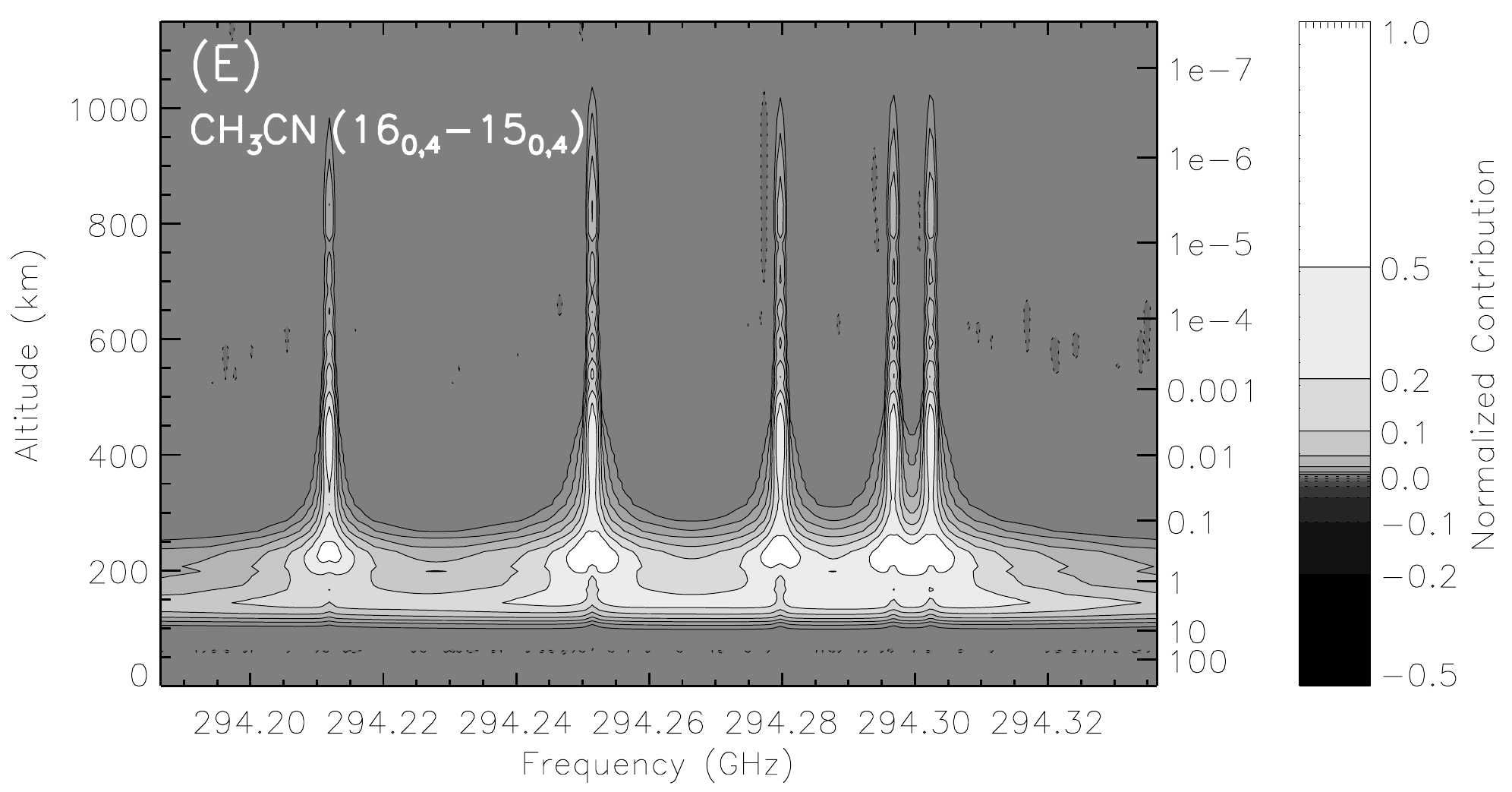}}
\end{center}
\caption{Contours of normalized functional
  derivatives \parencite{irwin_08} of spectral radiance per wavenumber
  with respect to chemical abundance for disk-averaged spectra of
  H$^{13}$CN (A), HC$^{15}$CN (B), HC$_3$N (C), C$_3$H$_4$ (D), and
  CH$_3$CN (E), as in Paper I and
  \textcite{molter_16}. Contour levels are 0, ± 0.0046, ± 0.01, ±
  0.0215, ± 0.046, ± 0.1, ± 0.215, and ± 0.46, and express molecular
  line sensitivity to volume mixing ratio at various pressure and altitude values.}
\label{fig:cf}
\end{figure}

\begin{figure}
\subfigure{\includegraphics[scale=0.95]{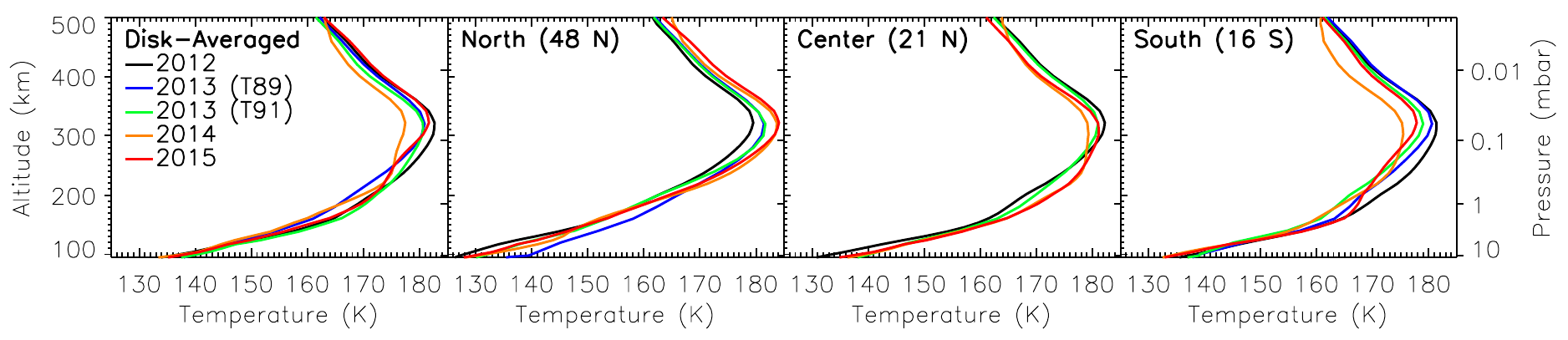}}
\caption{Temperature profiles from Paper I from 2012
    (black), 2014 (orange), and 2015 (red), and from the \textit{Cassini} T89 (blue)
    and T91 (green) flybys from \textcite{achterberg_14}.}
\label{fig:temps}
\end{figure}

\begin{figure}
\subfigure{\includegraphics[scale=1.0]{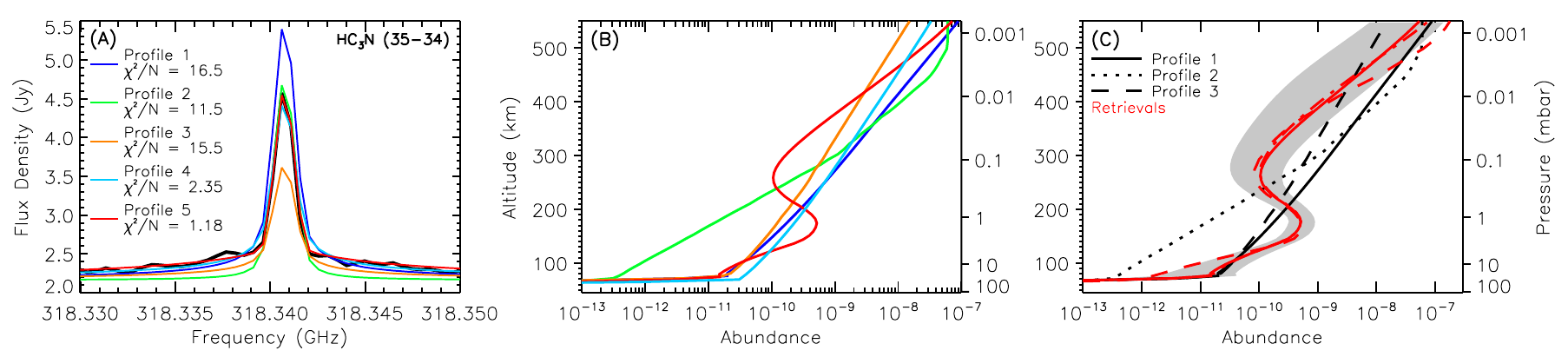}}
\caption{Retrieval tests for HC$_3$N disk-averaged spectrum from
  2014. (A) ALMA spectrum (black) and synthetic spectra for a variety
  of \textit{a priori} and retrieved profiles: 1, blue) linear
  gradient; 2, green)
  profile from \textcite{marten_02}; 3, orange) fractional scale height model from
  \textcite{cordiner_14}; 4, teal) fractional scale height retrieval;
  5, red)
  continuous retrieval. (B) Abundance profiles corresponding to
  spectra in A. (C) Comparison of retrieved profiles (red) using profiles
  1-3 as \textit{a priori} guesses (black). The retrieval errors for
  profile 1 (solid, red) are shown in gray.}
\label{fig:hc3n}
\end{figure}

\begin{figure}
\subfigure{\includegraphics[scale=0.9]{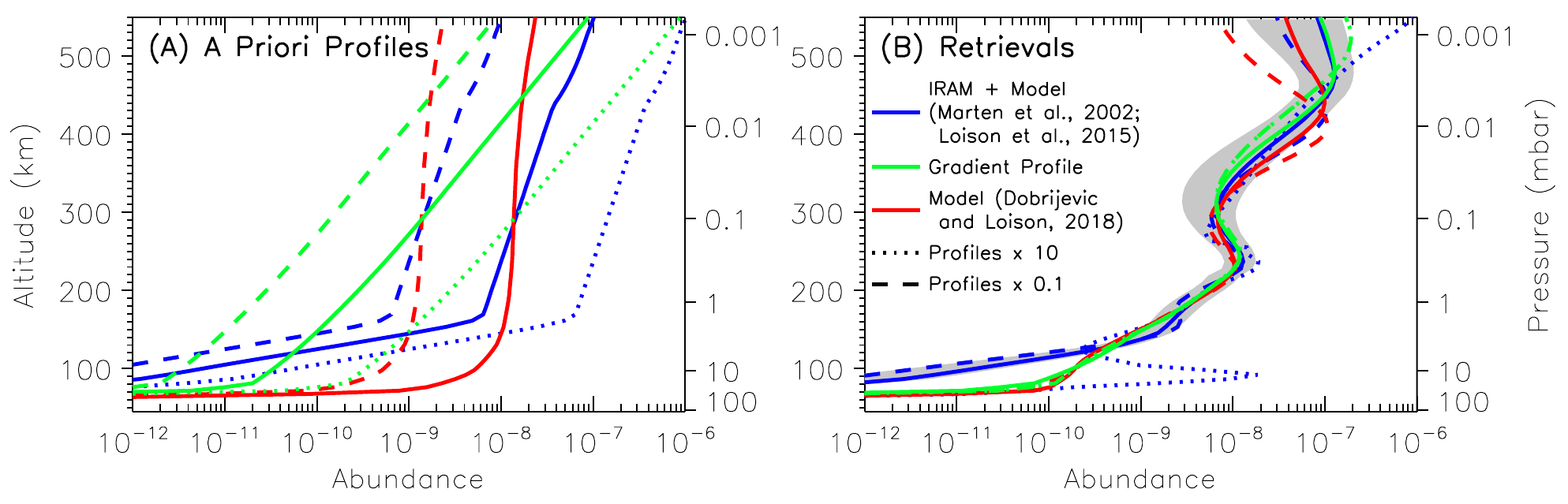}}
\caption{Retrieval tests for the CH$_3$CN disk-averaged spectrum from
  2014. (A) Plot of \textit{a priori} abundance profiles: (blue line) combination of sub-mm
  observations \parencite{marten_02} and photochemical
  model results \parencite{loison_15}; (red line)
  \textcite{dobrijevic_18} photochemical model; (green line) a test
  gradient profile. Dashed lines correspond to 10$\%$ of the solid line
  abundances; dotted lines are profiles with 10$\times$ the solid line
  abundances. (B) Retrieval results for each of the \textit{a priori}
  profiles in A. The error envelope for the solid blue retrieval
  (combination of \cite{marten_02} and \cite{loison_15}) is shown in
  gray. Retrieved profiles return to \textit{a priori} inputs above
  and below where the CH$_3$CN retrievals are sensitive
  ($\sim150-450$ km, see Fig. \ref{fig:cf}E).}
\label{fig:ch3cn}
\end{figure}

\begin{figure}
\subfigure{\includegraphics[scale=0.85]{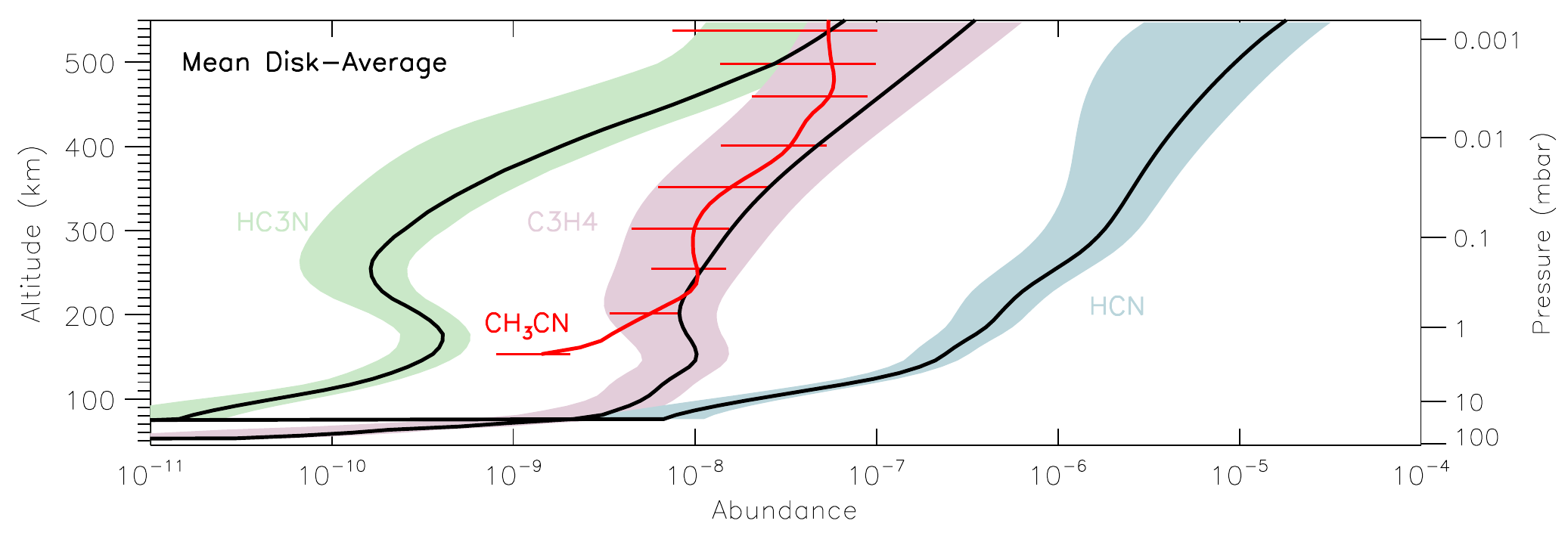}}
\caption{Disk-averaged abundance profiles found by taking the
  mean of measurements from each year. The HCN profile is an average
  of both scaled H$^{13}$CN and HC$^{15}$N retrievals. Retrieval
  errors are shown as shaded regions for HCN (blue), HC$_3$N (green),
  and C$_3$H$_4$ (lilac), and as bars for CH$_3$CN (red).}
\label{fig:abda}
\end{figure}

\begin{figure}
\begin{center}
\subfigure{\includegraphics[scale=1.]{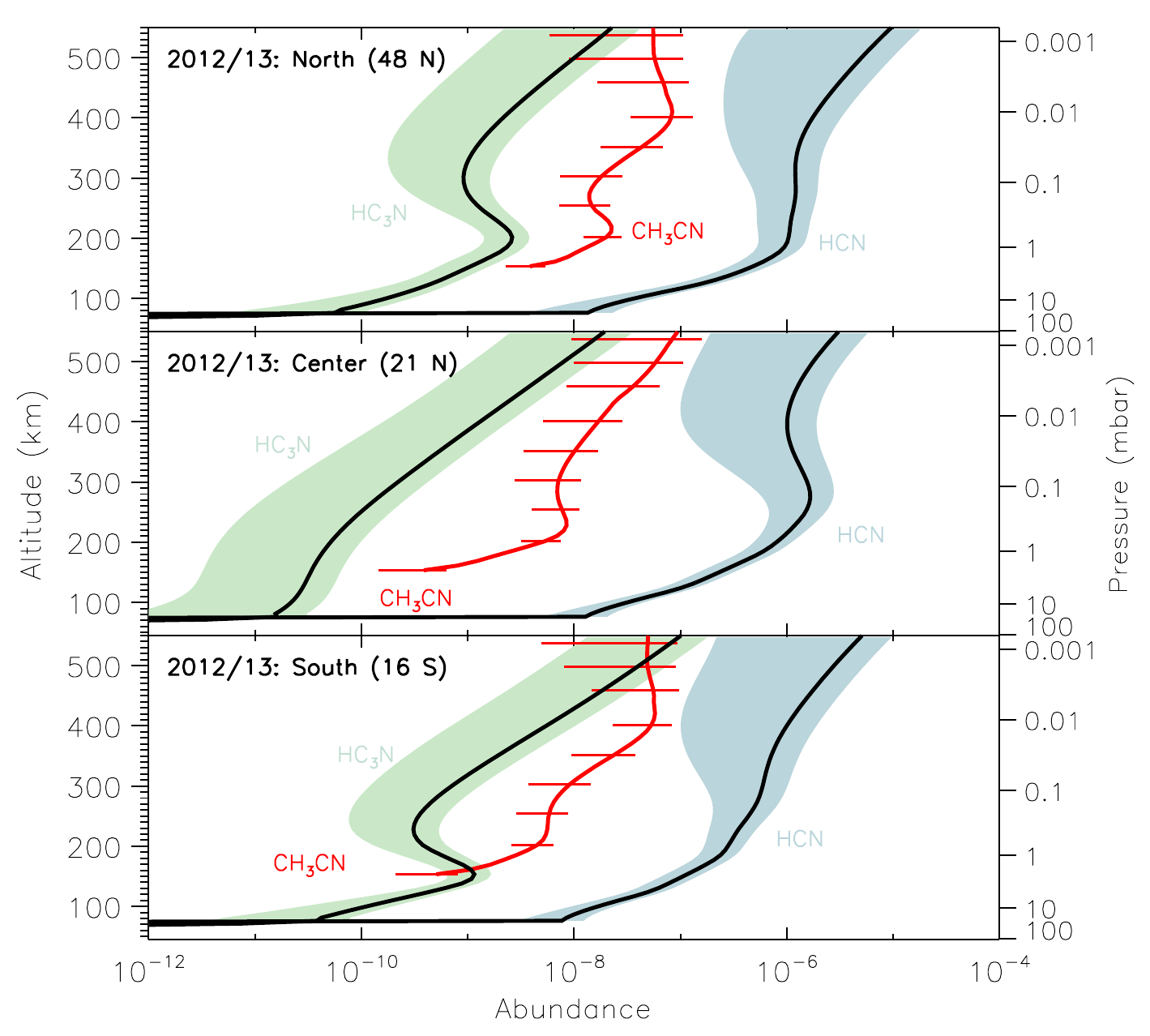}} 
\caption{Abundance profiles from retrievals of 2012 and 2013
  spectra in Fig. \ref{fig:spec_12} and \ref{fig:spec_13}. Retrieval
  errors are shown as shaded regions, except for bars corresponding to
  CH$_3$CN. HC$^{15}$N abundances have been scaled by the
  $^{14}$N/$^{15}$N ratio = 72.2 from \textcite{molter_16} to
  represent HCN here.}
\label{fig:ab12}
\end{center}
\end{figure}

\begin{figure}
\begin{center}
\subfigure{\includegraphics[scale=1.]{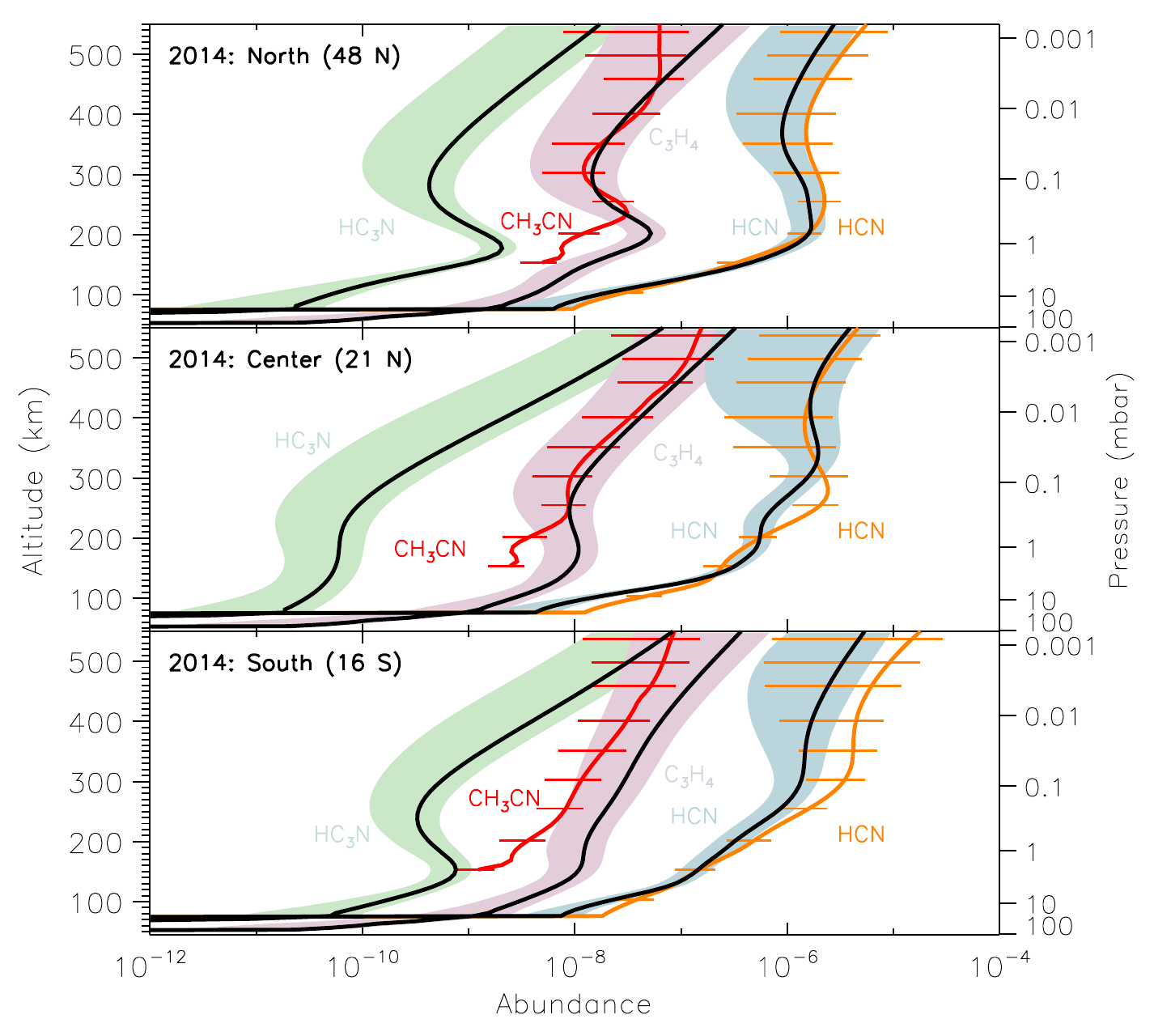}}
\caption{Abundance profiles for 2014 as in
  Fig. \ref{fig:ab12}. HCN abundances derived from HC$^{15}$N are shown in
  black with blue envelopes, and profiles
  derived from H$^{13}$CN are shown in orange with error bars, scaled by the $^{12}$C/$^{13}$C ratio = 89.8 from \textcite{molter_16}.}
\label{fig:ab14}
\end{center}
\end{figure}

\begin{figure}
\begin{center}
\subfigure{\includegraphics[scale=1.]{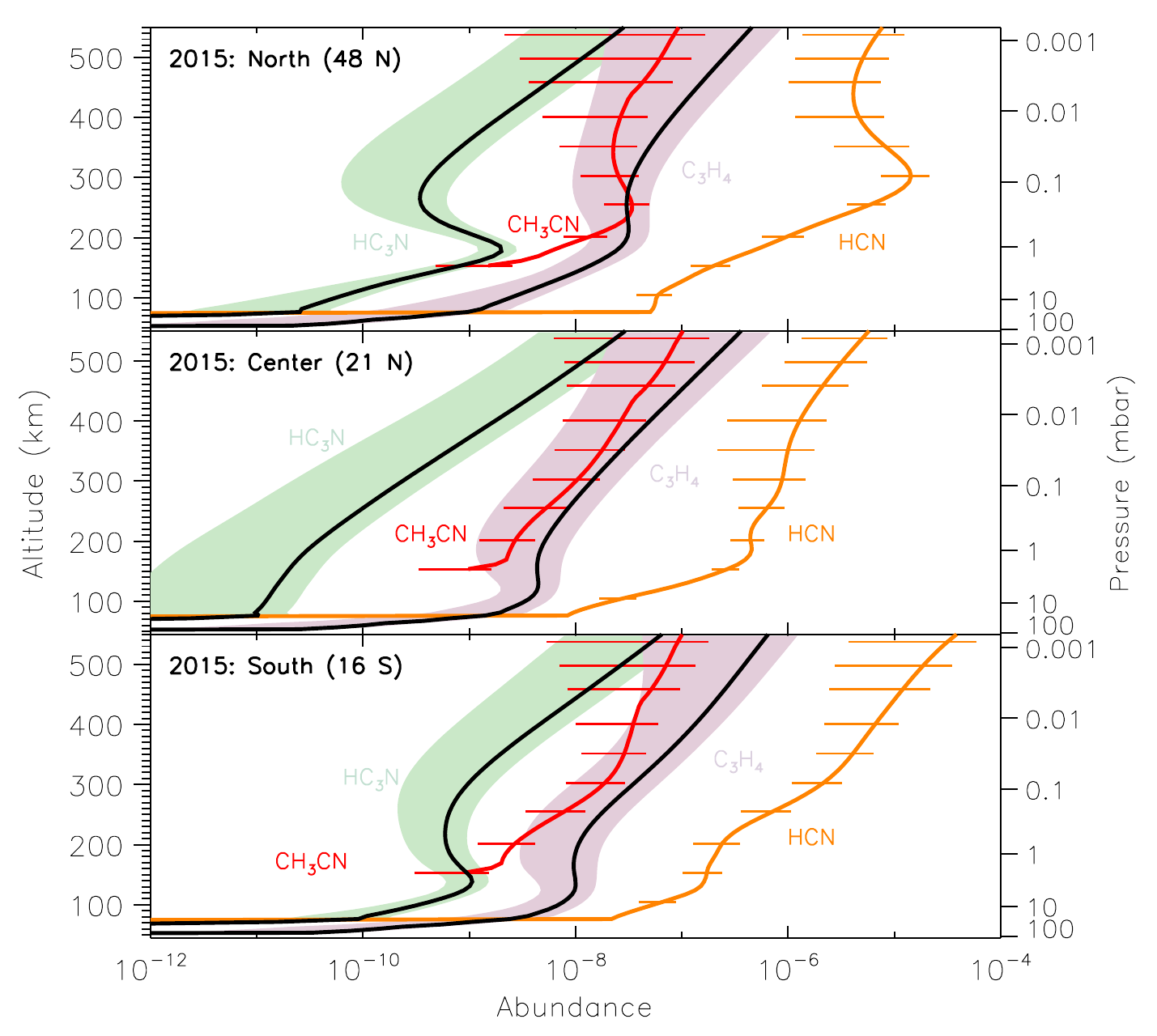}}
\caption{Abundance profiles for 2015, as in
  Fig. \ref{fig:ab12} and \ref{fig:ab14}.}
\label{fig:ab15}
\end{center}
\end{figure}

\begin{figure}
\subfigure{\includegraphics[scale=0.9]{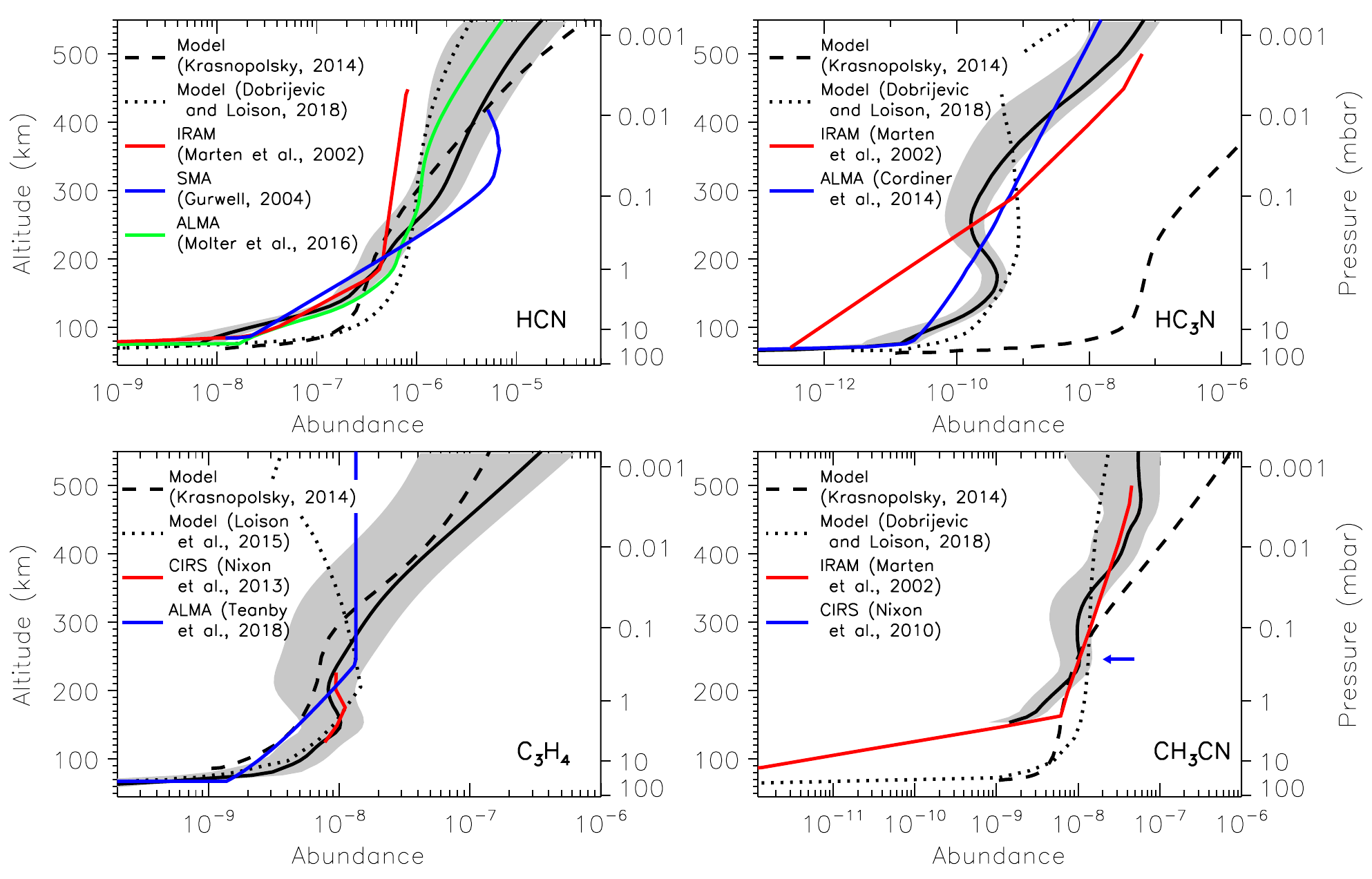}}
\caption{Comparisons of our mean, disk-averaged 
  abundance profiles (solid black lines) and the retrieval errors
  (gray envelopes) for each molecule (see Fig. \ref{fig:abda}) to: photochemical
  models (black dashed and dotted lines), including \textcite{krasnopolsky_14}, \textcite{loison_15},
  and \textcite{dobrijevic_18}; and retrieved profiles from various ground
  and space-based observatories (colored lines) -- including IRAM, the SMA, ALMA, and the
  \textit{Cassini} orbiter -- from \textcite{marten_02},
  \textcite{gurwell_04}, \textcite{nixon_10}, \textcite{nixon_13}, \textcite{cordiner_14},
  and \textcite{molter_16}.}
\label{fig:da_comp}
\end{figure}

\begin{figure}
\begin{center}
\subfigure{\includegraphics[scale=1.]{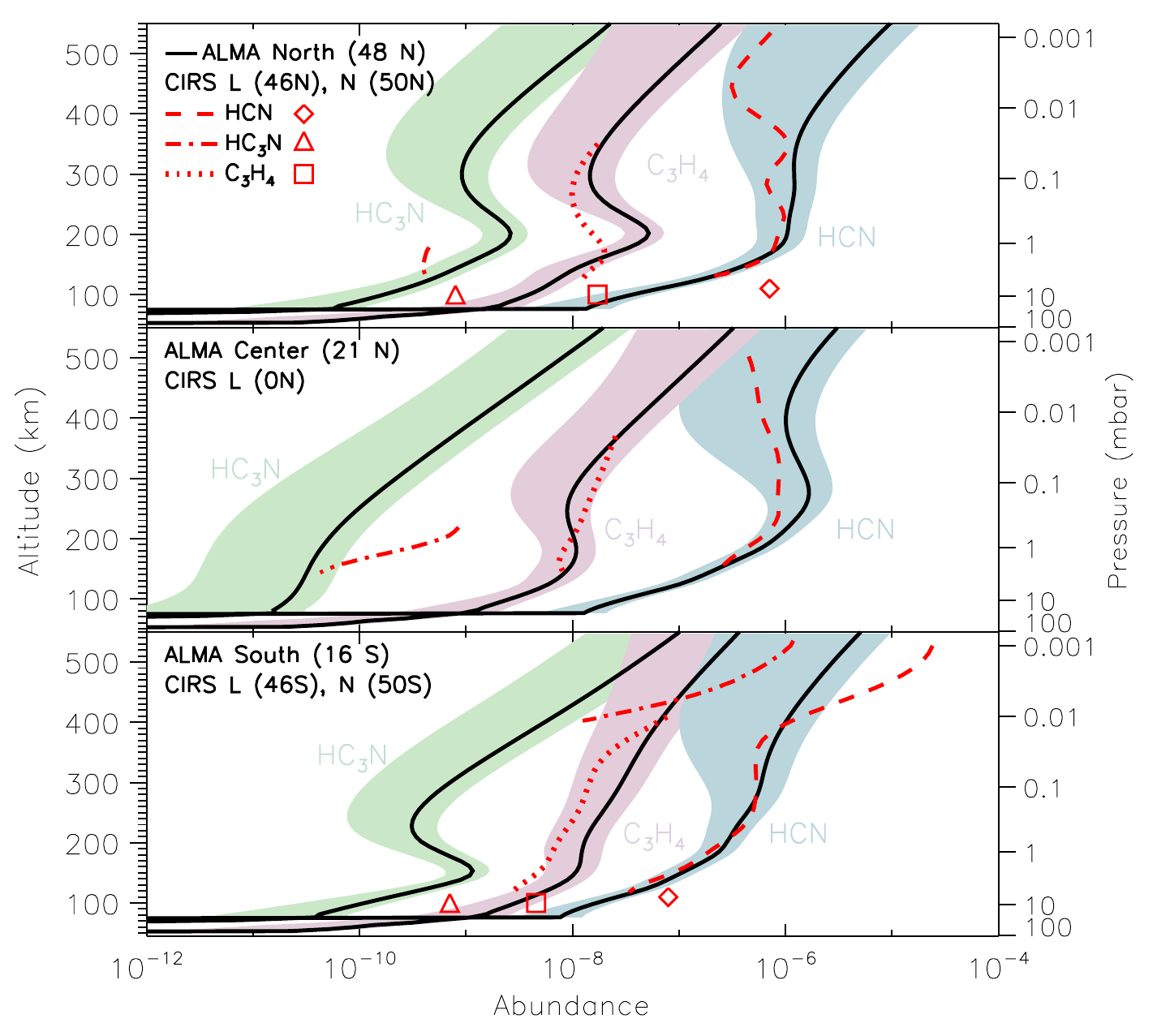}}
\caption{Comparisons of HCN, HC$_3$N profiles from
  Fig. \ref{fig:ab12} and C$_3$H$_4$ from Fig. \ref{fig:ab14} to
  \textit{Cassini}/CIRS limb (L; red lines) measurements from
  \textcite{vinatier_15} and nadir (N; red symbols)
  from measurements by \textcite{coustenis_16} at
  comparable latitudes.}
\label{fig:c_comp}
\end{center}
\end{figure}

\begin{figure}
\begin{center}
\subfigure{\includegraphics[scale=1.]{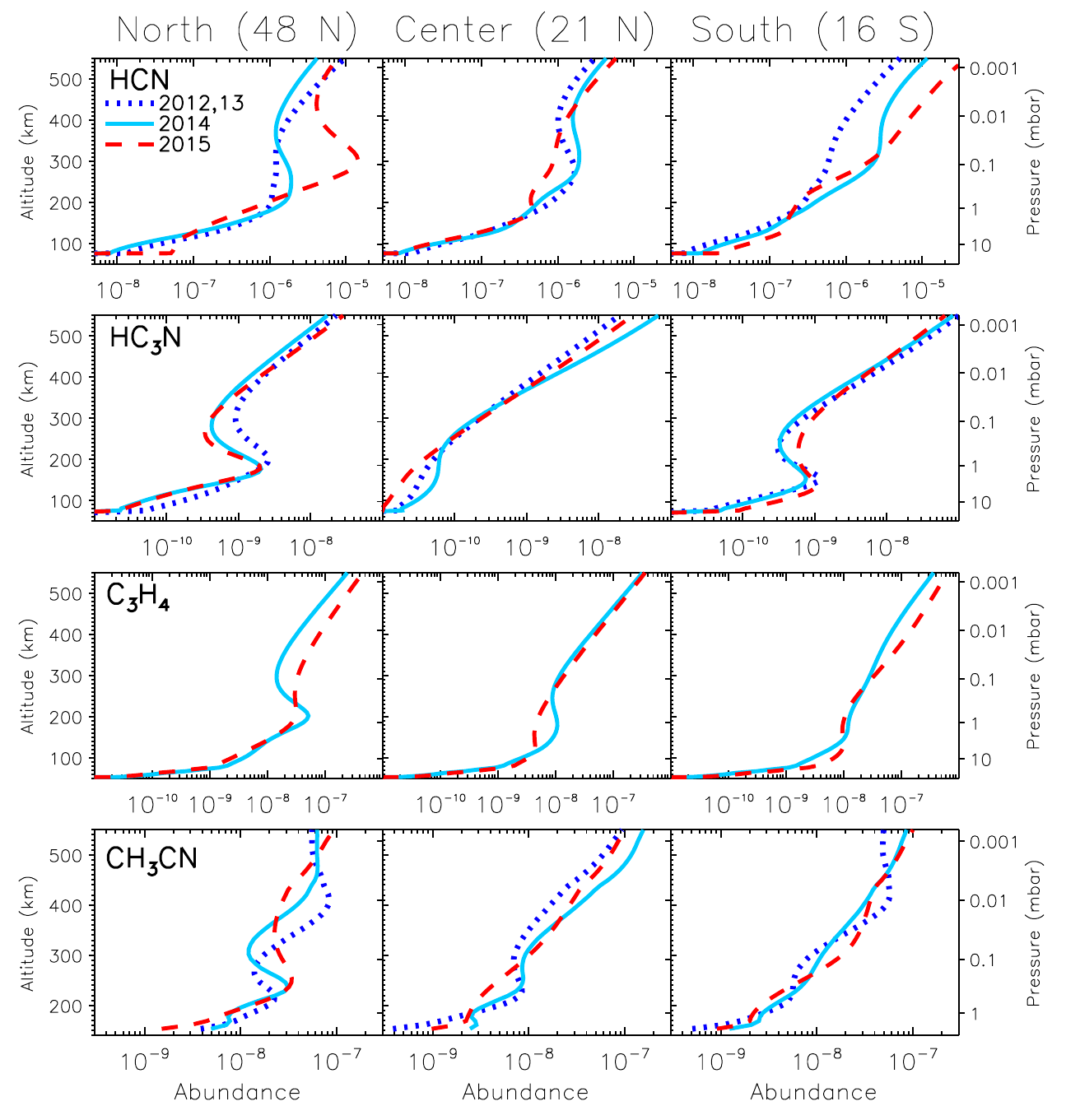}}
\caption{Temporal comparisons of abundance retrievals by species and
  region. North profiles in the first column; Center and South in
  the second and third columns. 2012
and 2013 retrievals are shown as blue dotted lines; 2014 profiles are shown in
solid teal lines; 2015 retrievals are shown as dashed red lines.}
\label{fig:temporal}
\end{center}
\end{figure}

While abundance comparisons to \textit{Cassini} are generally
in agreement, the HCN and HC$_3$N results show that our ALMA-derived retrievals
are missing the variability in vertical gradients and oscillations
seen at high latitudes on Titan due
to the spatial averaging of the relatively large ALMA beams, the inherent
vertical resolution constraints of ground-based (nadir) observations, and the
decreased sensitivity to altitudes $\textgreater300$ km. However,
our results still display large spatial variations in northern
and southern latitudes compared to the low northern (Center)
latitudes in both retrieved vertical profiles
(Fig. \ref{fig:ab12}--\ref{fig:ab15}) and intensity maps
(Fig. \ref{fig:maps}). When plotting profiles from each spatial region over time
(Fig. \ref{fig:temporal}), we can also see temporal trends arise as a
result of Titan's atmospheric dynamics, even at altitudes where ALMA
is less sensitive.

%HCN
The HCN isotopes that we model here lie on the broadened wings of CO
emission lines, making integrated flux maps difficult to
interpret. In the retrieval results, we observe a significant
enhancement in the north during 2015 near 0.1 mbar, which is $\sim$16 times greater than the abundance at
lower northern latitudes (Center) and a factor of 7 greater than the
south. A similar increase of HCN in mid-northern latitudes at similar
altitudes was observed in \textit{Cassini}/CIRS limb
data between 2011--2012 as the result of the of the weakening northern
polar vortex and the advection of accumulated enriched
gas to lower latitudes \parencite{vinatier_15}; the upper atmosphere (\textless0.01 mbar) in these
observations was also observed to be depleted in HCN, further
reinforcing the notion of a recent upwelling from the recent north-to-south
circulation cell. This trend is also present in our observations in
2014 and 2015, indicating we may be probing portions of the upper
atmosphere at higher northern latitudes that are now depleted in HCN
due to the rise of lower stratospheric air in the ascending
branch. Further, the retrieval results show a consistent enhancement of HCN in the
upper atmosphere ($\textgreater300$ km) of the low-southern latitudes over time
(Fig. \ref{fig:temporal}, top row), resulting in abundances
$\textgreater6$ times those of the Center and a factor of 5 greater
than the North. While our abundance retrievals are less sensitive to
emission at these altitudes, the trend in these profiles seems
significant; this trend is also observable at low
northern latitudes (Center) at the highest portion of our retrievals
(\textgreater400 km). Both of these increases are indicative of the circulation
of the large concentration of HCN formed at the north pole during
northern winter to the south (now winter) pole, and lower
latitudes. The effects of this new circulation are present in
contemporary \textit{Cassini}/CIRS measurements at high southern
latitudes \parencite{vinatier_15, coustenis_16, sylvestre_18}, where subsidence onto
the southern pole greatly increased the abundance of all species. In particular,  
species with long chemical lifetimes compared to dynamical timescales
in Titan's stratosphere (such as HCN; see e.g. \cite{loison_15}) provide good
tracers of Titan's global circulation \parencite{vinatier_15}. 

%HC3N
Retrievals of HC$_3$N for each year show significant enhancements in
the lower atmosphere in both the North and South profiles, and are the
largest spatial enhancements that we
measure here. While the North and South enhancements are reduced in 2014
-- with 34 and 13 times the Center abundance, respectively -- they
increase from a factor of 50 and 34 to 75 and 61 compared to the center from
2013 to 2015; these factors are larger than enhancements exhibited by
the other nitriles and C$_3$H$_4$ by an order of magnitude in the
lower stratosphere, but comparable to the large enrichment seen during
the northern winter by \textit{Cassini} \parencite{teanby_10b}. These
peaks most likely influence the large enhancement
seen in the disk-averaged profile from each year
(Fig. \ref{fig:abda}). A northern stratospheric enhancement of HC$_3$N may be the
result of advection between the polar vortex and lower latitudes, but
a `tongue' of enriched gas was not observed for HC$_3$N
during northern winter, as was seen for
HCN \parencite{teanby_08b}. A rapidly appearing tongue in the south
soon after the circulation reversal in 2011 also seems
unlikely (but motivates an analysis of the wind speeds during this
epoch). Further, lower atmosphere enhancements in abundance are
observed by photochemical models due to the influence of galactic
cosmic ray (GCR) induced chemistry; yet, the Center retrievals lack an
enhanced peak in the lower stratosphere, which would most likely manifest
regardless of latitude. This trend also does not fully agree with the integrated intensity
maps (Fig. \ref{fig:maps}), where the enhancements are only a
factor of 2--3 compared to the central flux, with a prominent decrease in the north from 2014 to
2015. The shift between a northern and southern enhancement of HC$_3$N
between 2014 and 2015 is consistent across ALMA observations of
Titan \parencite{cordiner_17a}. The discrepancy between retrieval
results and image maps may arise from the high opacity of the HC$_3$N line core in the
sub-mm, possibly inhibiting us from obtaining meaningful comparisons in the
lower atmosphere from integrated emission
maps \parencite{cordiner_18}. Finally, these spatial enhancements occur at
different altitudes -- near 150 km in the south and 200 km in the
north -- and the peak of both regions decreases by about 20 km from 2013 to
2015. The shift of these peaks with altitude and time may be a result
of a decrease in stratospheric temperatures, causing the
condensation altitude of HC$_3$N to change; this was observed in
\textit{Cassini}/CIRS
spectra at the south pole \parencite{jennings_12a,
  coustenis_16}. While ALMA temperature measurements at these same
spatial regions between 100--200 km reveal cooler temperatures at northern
latitudes compared to those from the subsolar point,
the temperatures at low southern latitudes are
comparable to those of the center (see Paper I, Fig. 9). 

HC$_3$N emission maps show significant spatial changes from
2013 to 2015, where we observe quickly increasing southern flux and decreasing flux in the
north, but these
large changes aren't immediately obvious in the retrieval results. We
do observe a general increase in southern abundances over time above and below
the abundance peak at $\sim150$ km, and a similar decrease in the north
(Fig. \ref{fig:temporal}, second row); we also observe a reduction in
abundance from Center retrievals \textgreater1 mbar from 2014 to 2015. Thus,
we can trace most of the variability in the integrated flux maps
to changes in the deeper atmosphere, and altitudes above the
potentially enriched reservoir of HC$_3$N near 1 mbar. The retrieved
profiles consistently show higher abundance in the upper atmosphere at
low southern latitudes by a factor of 2--4 compared to the North, but
the profiles do not show any significant trends over time. 

Our HC$_3$N North and South retrievals (and thus,
the disk-averaged results) may be adversely affected by high opacity and
large latitudinal
averages at low altitudes. As
found by \textcite{cordiner_18}, these effects may result in abundance
underestimates at the pole for higher spatial
resolution observations, but HC$_3$N still provides a valuable tracer
of meridional mixing of nitrile reservoirs from Titan's poles. If the 
enhancements we present here are real, the cause of a
lower stratospheric reservoir of HC$_3$N at mid to low latitudes is not
fully understood; this motivates a more
in depth study of HC$_3$N emission over time across Titan's limb,
where more accurate abundances may be derived at latitudes below
the poles.

%C3H4
As with the nitriles, we observe an enhancement of C$_3$H$_4$ at
mid-northern latitudes with factors of 5--6 greater than the Center
retrievals in 2014 and 2015. Similar to HCN and HC$_3$N, we find that the
Center C$_3$H$_4$ abundance decreases with time at altitudes
$\textless300$ km -- particularly from 2014 to 2015
(Fig. \ref{fig:temporal}, third row). We also observe a
simultaneous increase in the southern abundances of
C$_3$H$_4$ by a factor of 2 compared to the Center profiles, and a
general increase in the upper atmosphere
of both North and South retrievals. Both of these trends are
indicative of the redistribution of enriched gas from high northern
latitudes to the south with the reversal of Titan's circulation cell,
though we do not observe the increase in upper atmosphere gradient
observed with CIRS \parencite{vinatier_15}; this latter effect may be
missing from our C$_3$H$_4$ results due to the lack of sensitivity
above $\sim$400 km (Fig. \ref{fig:cf}D).
While both HC$_3$N and C$_3$H$_4$ did not show a significant lower
atmosphere tongue of enriched gas leaking from the polar vortex in
\textit{Cassini}/CIRS observations, C$_3$H$_4$ did extend further
past the vortex boundary than HC$_3$N by $10-15^\circ$ \parencite{teanby_09b}.
We also find that C$_3$H$_4$ is more enhanced at the Center compared to
HC$_3$N, as measured during northern winter. 

%CH3CN
CH$_3$CN shows enhancements in the north in both retrievals
(Fig. \ref{fig:temporal}, bottom row) and
maps (Fig. \ref{fig:maps}). In the latter, we see the emission peaks
confined to 45--60$^\circ$N and higher, consistent with some gas
advection beyond the northern polar
vortex barrier observed with \textit{Cassini} \parencite{teanby_08b}. We find a slight increase in lower
atmosphere abundances in the North over time, increasing by a factor of 3--6.5
compared to the Center near 1 mbar; this enhancement rises about 30 km
from 2013 to 2015. The decrease in northern emission seen in
the integrated flux maps may be an artifact caused by the increasing
spatial resolution over time, but we do observe a decrease in northern abundance
retrievals near 0.01 mbar (400 km) by a factor of $\sim$3 from 2013 to 2015,
where the contribution function for CH$_3$CN has a secondary peak
(Fig. \ref{fig:cf}E). This change is minor
compared to the retrieval errors (i.e. $\textless2\sigma$), but could be the result of upwelling
of depleted air from the lower atmosphere that we see with HCN and as
observed by CIRS \parencite{vinatier_15}. The
enhancement of CH$_3$CN in the northern lower atmosphere 
is indicative of a winter enrichment of this molecule, which may be
advected to the lower latitudes after northern winter. CH$_3$CN
has a relatively long chemical lifetime throughout 
Titan's atmosphere (as compared to the dynamical
lifetime), with a similar lifetime to HCN in the stratosphere
\parencite{wilson_04, loison_15}. However, we
don't find large variations in southern abundances at higher
altitudes over time, or a large difference between North and South
abundances in the upper stratosphere as is observed for HCN here. As with the other nitriles,
observations of these changes at the southern pole
are inhibited by our viewing angle from Earth, though the emission
maps may provide evidence for circulation from the north pole to the south over time.

Vertical oscillations appear in CH$_3$CN retrievals to a larger extent
than the other molecules, particularly in 2014. Oscillations in
previous \textit{Cassini} measurements of nitriles have been
documented, particularly for mid to high northern latitudes, and are thought
to be the result of small scale dynamical mixing between gas-depleted lower latitudes and
the enriched polar vortex \parencite{teanby_09c}. While our CH$_3$CN retrievals exhibit
larger vertical oscillations with increasing northern latitudes (Fig. \ref{fig:temporal}), we do
not see a similar pattern in the other nitriles or C$_3$H$_4$, as were
seen in \textit{Cassini}/CIRS results \parencite{teanby_09c,
  vinatier_15}. The lack of contemporary, spatially resolved abundance
measurements for CH$_3$CN in Titan's stratosphere, combined with the
averaging of our measurements across multiple latitudes (with
potentially significant dynamics and meridional mixing) makes these vertical
oscillations difficult to interpret.

\section{Conclusions}
Building on the previous results in Paper I, we present vertical abundance profiles of HCN, HC$_3$N, C$_3$H$_4$,
and CH$_3$CN obtained through the analysis
of rotational transitions in spatially
resolved (beam sizes $\sim0.2-0.5''$) ALMA flux calibration data from 2012 to 2015.
The comparison of three regions on Titan's disk (centered at $\sim$48$^\circ$N, 21$^\circ$N,
and 16$^\circ$S) reveal distinct spatial variations and insight
into Titan's atmospheric dynamics. In contrast with the temperature profiles presented in Paper I, the
abundance profiles of these molecules show temporal changes over the 3 years of
observation from 2012 to 2015. The combination of the spatial and
temporal variations we observe informs our understanding of Titan's
atmospheric circulation into northern spring and summer. Our findings
are summarized as follows:
\begin{itemize}
 \item All four molecules display enhancements in the North (and often
   the South) compared to Center, ranging from factors of $\sim$6 in C$_3$H$_4$ and CH$_3$CN to 15 and 75 in
   HCN and HC$_3$N, respectively. Southern enhancements are more noticeable in the upper
   atmosphere, particularly for HCN and HC$_3$N, yet do not exhibit the steep
   vertical gradients seen by
   \textit{Cassini}/CIRS \parencite{teanby_12, vinatier_15, coustenis_16}.
 \item We find large enhancements of HC$_3$N between 150--200 km in
   all North and South retrievals. This is indicative of a relatively
   new lower atmosphere HC$_3$N reservoir, but may also be the result of
   opacity effects of sub-mm HC$_3$N
   lines \parencite{cordiner_18}. Nevertheless, the combination of retrieval
   results and integrated flux maps show a rapid reduction of HC$_3$N
   at Titan's north pole and a simultaneous increase in the south
   between 2013 and 2015.
 \item We observe many temporal trends in abundance retrievals that
   reveal the continued effects of Titan's large north-to-south
   circulation cell:
  \begin{enumerate}
    \item The increase of southern HCN, HC$_3$N, C$_3$H$_4$, and
      potentially CH$_3$CN at higher altitudes, and
      similar trends (with reduced magnitude) at low-northern latitudes (Center).
    \item A slight increase in the abundances of mid northern latitudes over time
      in HCN, C$_3$H$_4$, CH$_3$CN, with a change in upper atmosphere
      gradients in the longer lived chemical species (HCN and
      CH$_3$CN).
    \item An increase in abundance for all molecules at pressures \textgreater1
      mbar at southern latitudes, except CH$_3$CN, which does not
      effectively sound higher pressures.
    \item A reduction in abundance for all molecules in Center
      profiles at pressures \textgreater0.1 mbar.
  \end{enumerate}
  These trends show evidence for subsidence at the southern pole, 
  the decrease of a `tongue' at low northern latitudes
  (where enriched air was advected from the northern polar vortex during winter), and lofted
  air replete with longer lived chemical species from Titan's lower
  stratosphere to higher altitudes.
 \item The polar enhancements and vertical gradients observed are
  generally less significant than those observed with the
  \textit{Cassini} orbiter, indicating that the effects of
  atmospheric chemistry and dynamics are muted when observed in large
  latitudinal averages (as seen with the temperatures reported in
  Paper I).
\end{itemize}
We validated our results using
contemporaneous \textit{Cassini}/CIRS data, and through comparisons of mean
profiles from 2012 to 2015 to previous
disk-averaged ground-based observations and photochemical model
results. We find that our retrieved profiles are comparable to contemporary
studies with the exception of HC$_3$N, which is optically thick at the
poles where the molecule has been observed to be greatly
enhanced. However, the large temporal variations and fine vertical
structure observed with the \textit{Cassini} orbiter are obscured by
the latitudinal averaged measurements derived from spectra
representing relatively large ALMA beam-footprints, particularly at
higher altitudes where sub-mm measurements are not as sensitive, as observed
with the previous atmospheric temperature retrievals (Paper I). Thus,
this work serves as a proof of concept for future measurements
of Titan's chemical abundances throughout the stratosphere
that will allow us to continue
monitoring Titan's varied atmospheric dynamics into the post-\textit{Cassini} era.

\section{Acknowledgments}
This research was supported by NASA’s Office of Education and the NASA
Minority University Research and Education Project ASTAR/JGFP Grant
$\#$NNX15AU59H. Additional funding was provided by the NRAO Student
Observing Support award $\#$SOSPA3-012.
%CO-I funding:
C.A.N was supported in this work by the
NASA Solar System Observations Program and the NASA Astrobiology Institute.
M.A.C received funding from the National Science Foundation under Grant
No. AST-1616306. N.A.T and P.G.J.I were supported by the UK Science
and Technology Facilities Council. S.B.C was funded by an award from the NASA Science
Innovation Fund. 
%ALMA
This paper makes use of the following ALMA data:
ADS/JAO.ALMA$\#$2011.0.00724.S, 2011.0.00820.S, 2012.1.00377.S,
2012.1.00225.S, 2012.1.00453.S, 2013.1.00220.S, and 2013.1.00111.S. ALMA is a partnership of ESO (representing its member
states), NSF (USA) and NINS (Japan), together with NRC (Canada) and
NSC and ASIAA (Taiwan) and KASI (Republic of Korea), in cooperation
with the Republic of Chile. The Joint ALMA Observatory is operated by
ESO, AUI/NRAO and NAOJ. The National Radio Astronomy Observatory is a
facility of the National Science Foundation operated under cooperative
agreement by Associated Universities, Inc.

The authors would like to thank Richard Achterberg for his
insightful comments on Titan's stratospheric temperatures and his
contribution of \textit{Cassini}/CIRS temperature profiles for this work
and Paper I.

%\end{linenumbers}

%%%%% References %%%%%
\printbibliography[title={References}]

%%%%% Tables and Figures %%%%%
\newpage

%---------------Observation Parameters Table-------------------
%\singlespacing
\begin{table}\scriptsize
\begin{center}
\caption[]{Observational Parameters}
\begin{tabular}{ccccccc}
\toprule
\textbf{Project} & \textbf{Observation} & \textbf{$\#$ of} &
                                                             \textbf{Integration}
  & \textbf{Spectral} & \textbf{Beam} & \textbf{Species} \\
 \textbf{ID} & \textbf{Date} & \textbf{Antennas} & \textbf{Time (s)} & 
                                                                     \textbf{Res. (kHz)} & \textbf{Size$^a$} & \\
\midrule
\midrule
\textbf{2012} \\
\midrule
2011.0.00724.S & 05 Jun 2012 & 21 & 236 & 976 & 0.32$''$ $\times$
                                                0.25$''$ & HC$^{15}$N \\
\midrule
\midrule
\textbf{2013} \\
\midrule
2011.0.00820.S & 01 Jan 2013 & 24 & 66 & 976 &0.66$''$
                                                      $\times$
                                                      0.53$''$& C$_3$H$_4$ \\
2012.1.00377.S & 01 Jun 2013 & 30 & 157 & 976 &
 0.65$''$ $\times$ 0.33$''$* & HC$_3$N \\
 & & & & & 0.50$''$ $\times$ 0.35$''$* & CH$_3$CN\\
\midrule
\midrule
\textbf{2014} \\
\midrule
2012.1.00225.S & 14 Apr 2014 & 34 & 158 & 976 & 0.29$''$ $\times$ 0.20$''$& H$^{13}$CN \\
2012.1.00453.S & 28 Apr 2014 & 35 & 158 & 976 & 0.37$''$ $\times$
                                                0.30$''$* & HC$^{15}$N \\
 & 08 Jul 2014 & 31 & 157 & 976 & 0.45$''$ $\times$ 0.36$''$ & CH$_3$CN \\ 
 & 16 Jul 2014 & 32 & 157 & 976 & 0.36$''$ $\times$ 0.35$''$ & HC$_3$N \\ 
& & & & & 0.38$''$ $\times$ 0.36$''$ & C$_3$H$_4$ \\
\midrule
\midrule
\textbf{2015} \\
\midrule
2012.1.00377.S & 19 May 2015 & 37 & 157 & 976 & 0.26$''$ $\times$ 0.22$''$& H$^{13}$CN\\
 & & & & & 0.25$''$ $\times$ 0.24$''$ & CH$_3$CN \\
 2013.1.00220.S & 14 Jun 2015 & 41 & 157 & 1953 &
        0.38$''$ $\times$ 0.36$''$* & C$_3$H$_4$, C$_2$H$_5$CN \\
2013.1.00111.S & 22 Jul 2015 & 44 & 261 & 976 & 0.35$''$ $\times$ 0.30$''$* & HC$_3$N \\
\bottomrule
\label{tab:obs}
\end{tabular}
\end{center}
{\scriptsize{{\bf{Notes:}} $^a$FWHM of the Gaussian restoring
    beam. *Denotes resolution obtained through Briggs weighting as
    opposed to Natural.}}
\end{table}
%\doublespacing
%----------------------Table End --------------------------

%Integrated Flux Maps
%----------------------------------------------------
%----------------------------------------------------

%-----------------Spectral Transitions--------------------
%\singlespacing
\begin{table}\scriptsize
\begin{center}
\caption[]{Spectral Transitions}
\begin{tabular}{ccc}
\toprule
\textbf{Species} & \textbf{Transition} & \textbf{Rest Freq.}\\
& \textbf{($J''_{K''_{a,c}}-J'_{K'_{a,c}}$)} & \textbf{(GHz)}\\
\midrule
\midrule
\textbf{2012} \\
\midrule
HC$^{15}$N & 8--7 & 688.273 \\
\midrule
\midrule
\textbf{2013} \\
\midrule
C$_3$H$_4$ & 15$_3$--14$_3$ & 256.293 \\
C$_3$H$_4$ & 15$_2$--14$_2$ & 256.317 \\
C$_3$H$_4$ & 15$_1$--14$_1$ & 256.332 \\
C$_3$H$_4$ & 15$_0$--14$_0$ & 256.337 \\
\\
HC$_3$N & 40--39 & 363.785 \\
\\
%CH$_3$CN & 19$_6$--18$_{-6}$ & 349.212 \\
CH$_3$CN & 19$_6$--18$_{6}$ & 349.212 \\
CH$_3$CN & 19$_5$--18$_{5}$ & 349.286 \\
CH$_3$CN & 19$_4$--18$_{4}$ & 349.346 \\
CH$_3$CN & 19$_3$--18$_{3}$ & 349.393\\
%CH$_3$CN & 19$_3$--18$_{-3}$ & 349.393\\
%CH$_3$CN & 19$_{-3}$--18$_{3}$ & 349.393 \\
CH$_3$CN & 19$_2$--18$_{2}$ & 349.426 \\
CH$_3$CN & 19$_1$--18$_{1}$ & 349.446 \\
CH$_3$CN & 19$_0$--18$_{0}$ & 349.453 \\
\midrule
\midrule
\textbf{2014} \\
\midrule
H$^{13}$CN & 8--7 & 690.552 \\
\\
HC$^{15}$N & 4--3 & 344.200 \\
\\
CH$_3$CN & 16$_4$--15$_4$ & 294.212 \\ 
CH$_3$CN & 16$_3$--15$_{3}$ & 294.251 \\ 
%CH$_3$CN & 16$_3$--15$_{-3}$ & 294.251 \\ 
%CH$_3$CN & 16$_{-3}$--15$_3$ & 294.251 \\ 
CH$_3$CN & 16$_2$--15$_2$ & 294.280 \\ 
CH$_3$CN & 16$_1$--15$_1$ & 294.297 \\ 
CH$_3$CN & 16$_0$--15$_0$ & 294.302  \\ 
\\
HC$_3$N & 35--34 & 318.341 \\ 
\\
C$_3$H$_4$ & 18$_4$--17$_4$ & 307.489 \\
C$_3$H$_4$ & 18$_3$--17$_3$ & 307.530\\
C$_3$H$_4$ & 18$_2$--17$_2$ & 307.560 \\
C$_3$H$_4$ & 18$_1$--17$_1$ & 307.577  \\
C$_3$H$_4$ & 18$_0$--17$_0$ & 307.583  \\
\midrule
\midrule
\textbf{2015} \\
\midrule
H$^{13}$CN & 8--7 & 690.552 \\
\\
CH$_3$CN & 37$_{3}$--36$_3$ & 679.831 \\
%CH$_3$CN & 37$_{-3}$--36$_3$ & 679.831 \\
%CH$_3$CN & 37$_3$--36$_{-3}$ & 679.831 \\
CH$_3$CN & 37$_2$--36$_2$ & 679.895 \\
CH$_3$CN & 37$_1$--36$_1$ & 679.934 \\
CH$_3$CN & 37$_0$--36$_0$ & 679.947 \\
\\
C$_3$H$_4$ & 20$_3$--19$_3$ & 341.682 \\
C$_3$H$_4$ & 20$_2$--19$_2$ & 341.715 \\
C$_3$H$_4$ & 20$_1$--19$_1$ & 341.735 \\
C$_3$H$_4$ & 20$_0$--19$_0$ & 341.741 \\
\\
C$_2$H$_5$CN & 40$_{1,40}$--39$_{1,39}$ & 341.704\\
C$_2$H$_5$CN & 40$_{0,40}$--39$_{0,39}$ & 341.711\\
\\
HC$_3$N  & 24--23 & 218.325 \\
\bottomrule
\label{tab:lines}
\end{tabular}
\end{center}
\end{table}
%\doublespacing
%----------------------Table End --------------------------

%Spectra plot 2012
%----------------------------------------------------
%----------------------------------------------------

%Spectra plot 2013
%----------------------------------------------------
%----------------------------------------------------

%Spectra plot 2014
%----------------------------------------------------
%----------------------------------------------------

%Spectra plot 2015
%----------------------------------------------------
%----------------------------------------------------

%Contribution Functions
%----------------------------------------------------
%----------------------------------------------------

%Temperature plot
%----------------------------------------------------
%----------------------------------------------------

%HC3N test plot
%----------------------------------------------------
%----------------------------------------------------

%CH3CN test plot
%----------------------------------------------------
%----------------------------------------------------

%Disk-Averaged Abundances
%----------------------------------------------------
%----------------------------------------------------

%2012-13 Abundances
%----------------------------------------------------
%----------------------------------------------------

%2014 Abundances
%----------------------------------------------------
%----------------------------------------------------

%2015 Abundances
%----------------------------------------------------
%----------------------------------------------------

%Disk-averaged comparisons
%----------------------------------------------------
%----------------------------------------------------

%Cassini Comparisons
%----------------------------------------------------
%----------------------------------------------------

%Temporal Variations
%----------------------------------------------------
%----------------------------------------------------

%%%%% End of document %%%%%

\end{document}